\documentclass{article}
\usepackage{slashed,verbatim,subfigure}
\usepackage[numbers,sort&compress]{natbib}
\usepackage{amsmath}
\usepackage{amssymb}
\usepackage{appendix}
\usepackage{graphicx}
\usepackage{dcolumn}
\usepackage{authblk}
\usepackage{bm}
\usepackage[colorlinks=true,linktocpage=true,
linkcolor=blue,citecolor=blue]{hyperref}
\usepackage[a4paper]{geometry}
\newcommand{\be}{\begin{eqnarray}}
\newcommand{\ee}{\end{eqnarray}}

\begin{document}
\large
\title{\bf{Transport phenomena and observables associated with viscous properties of an anisotropic hot QCD medium at finite baryon asymmetry}}
\author{Shubhalaxmi Rath\thanks{shubhalaxmirath@gmail.com} }
\author{Nicol\'{a}s A. Neill\thanks{naneill@outlook.com}}
\affil{Centro Multidisciplinario de F\'isica, Vicerrector\'ia de Investigaci\'on, Universidad Mayor, 8580745 Santiago, Chile}
\date{}
\maketitle

\begin{abstract}
We have studied the shear viscosity, bulk viscosity and associated observables of baryon asymmetric hot QCD medium subjected to weak momentum anisotropy induced by the asymptotic expansion of the medium 
during the early stages of ultrarelativistic heavy-ion collisions. The shear and bulk viscosities are evaluated by solving the relativistic Boltzmann transport equation within the relaxation time approximation in a kinetic theory framework. We have further analysed the Prandtl number, the Reynolds number, specific shear viscosity, and specific bulk viscosity, characterizing the sound attenuation, flow properties, fluidity, and conformal behavior, respectively. The interactions among the constituent partons are incorporated through the anisotropic quasiparticle distribution functions, with the medium characterized by finite temperature, momentum anisotropy and baryon asymmetry. Our analysis shows that both shear and bulk viscosities decrease with increasing anisotropy in both baryonless and baryon asymmetric media. In contrast, finite baryon asymmetry enhances both viscosities compared with their baryonless counterparts. The effects of expansion-induced anisotropy remain pronounced in both the baryonless and baryon asymmetric media, leading to substantial modifications of the associated viscous observables. These features are attributed primarily to the squeezing of the parton distribution functions caused by expansion-induced momentum anisotropy, together with the resulting modification of the partonic dispersion relations. Our results demonstrate that the expansion-induced anisotropy significantly influences the sound attenuation, flow characteristics, fluidity and conformal behavior of baryon asymmetric hot QCD medium, providing new insights into the viscous properties of the medium created in ultrarelativistic heavy-ion collisions. 

\end{abstract}

\newpage

\section{Introduction}
The high energy densities reached in ultrarelativistic heavy-ion collisions at the Relativistic Heavy Ion Collider (RHIC) and the Large Hadron Collider (LHC) provide a unique environment for producing quark-gluon plasma (QGP), an extreme phase of the strongly interacting QCD matter. The properties of this extreme phase are influenced by several conditions established during the early stages of the collisions. One such condition is momentum anisotropy, which develops in the local rest frame of the fireball due to the faster expansion along the beam (longitudinal) direction compared to the transverse plane \cite{Dumitru:PLB662'2008,Dumitru:PRD79'2009}. This anisotropy can be characterized by a parameter ($\xi$) defined in terms of the transverse ($p_T$) and longitudinal ($p_L$) momentum components. In the weak-anisotropy limit ($\xi<1$), the parton distribution function may be approximated as a deformation of an isotropic distribution, effectively compressed along a preferred direction. Such modifications alter the microscopic phase-space structure of the medium and, consequently, its transport properties. Within the framework of kinetic theory, the transport coefficients are determined by the distribution functions and dispersion relations of partons. The introduction of anisotropy modifies these inputs, leading to corresponding changes in the transport coefficients. Recent studies have highlighted the importance of explicitly incorporating anisotropy into the transport modeling to obtain realistic descriptions of the QGP produced in heavy-ion collisions \cite{Martinez:PRC78'2008,Thakur:PRD88'2013,Ryblewski:PRD92'2015,Bhattacharya:PRD93'2016,
Mukherjee:EPJA53'2017,Rath:PRD100'2019,Rath:PRD102'2020,Rath:EPJA62'2026,Rath:PRD113'2026}. Moreover, the early stages of these collisions may exhibit a small but finite baryon asymmetry. For instance, at temperatures around 160 MeV, the baryon chemical potential is estimated to be of the order of 300 MeV \cite{P:JPG28'2002,Cleymans:JPG35'2008,Andronic:NPA837'2010}. In the presence of a strong magnetic field, this value can increase significantly \cite{Fukushima:PRL117'2016}, implying a corresponding rise in the quark chemical potential. The combined effects of magnetic fields and baryon asymmetry have been shown to substantially influence various transport coefficients of the partonic medium \cite{Rath:EPJC80'2020,Rath:EPJC81'2021,Rath:EPJC82'2022,Rath:EPJA59'2023}. Additionally, in chiral systems at finite temperature, baryon asymmetry can induce axial currents \cite{Pu:PRD91'2015,Gorbar:PRD93'2016}. Motivated by these considerations, it is essential to investigate transport coefficients in a more realistic setting that simultaneously accounts for momentum-space anisotropy and baryon asymmetry. In this work, we examine the impact of expansion-induced anisotropy on shear viscosity, bulk viscosity and related observables of a baryon asymmetric hot QCD medium. 

Momentum transport coefficients, particularly the shear viscosity ($\eta$) and the bulk viscosity ($\zeta$), play a central role in the hydrodynamic description of the QGP. Shear viscosity characterizes the transfer of momentum within the medium, whereas bulk viscosity quantifies the change in local pressure associated with expansion or compression. In this sense, shear viscosity measures the resistance to deformation at constant volume, while bulk viscosity reflects the resistance to volume changes at fixed shape. These coefficients significantly influence experimental observables in heavy-ion collisions, such as the elliptic flow coefficient and the transverse momentum spectra of produced hadrons. The ratio of shear viscosity to entropy density ($\eta/s$) provides key insight into the fluidity of the medium, whereas the ratio of bulk viscosity to entropy density ($\zeta/s$) serves as an indicator of conformal symmetry. Both ratios are also important for identifying the phase transition of the QGP. An accurate determination of $\eta$ and $\zeta$ is therefore important for establishing a quantitative connection between the viscous properties of the QGP and experimentally measured observables such as particle multiplicities, transverse momentum spectra and elliptic flow \cite{Ryu:PRL115'2015}. According to the Anti-de Sitter/Conformal Field Theory (AdS/CFT) correspondence, $\eta/s$ has a lower bound at $1/(4\pi)$, providing an important benchmark for characterizing the QGP as a strongly coupled nearly perfect fluid \cite{Kovtun:PRL94'2005}, whereas perturbative QCD calculations yield substantially larger values of $\eta/s$ \cite{Arnold:PRD74'2006}. In contrast, $\zeta/s$ tends to be very small and vanishes in conformally symmetric systems. However, lattice QCD results \cite{Borsonyi:JHEP11'2010,Bazavov:PRD90'2014} reveal a pronounced peak in the trace anomaly near the critical temperature, demonstrating that QCD deviates from conformal symmetry in this region. External conditions can further modify these transport coefficients. For instance, the strong magnetic fields introduce anisotropies such that the shear and bulk viscosities develop components along the magnetic field direction, particularly in the strong magnetic field regime \cite{Lifshitz:BOOK'1981,Tuchin:JPG39'2012,Hattori:PRD96'2017}. Similarly, rapid rotation has been shown to significantly affect their magnitudes \cite{Rath:PRD112'2025}. In addition, expansion-induced momentum anisotropy can substantially alter these transport properties of the hot QCD medium. A recent study indicates that such anisotropy has a marked impact on charge and heat transport coefficients \cite{Rath:EPJA62'2026}, motivating a detailed study of its effect on momentum transport coefficients as well. In this work, we systematically examine how expansion-induced anisotropy influences shear and bulk viscosities in a baryon asymmetric hot QCD medium. Furthermore, we analyze the combined effects of baryon asymmetry and anisotropy on related observables, including the Prandtl number (Pr), the Reynolds number (Re), the specific shear viscosity ($\eta/s$) and the specific bulk viscosity ($\zeta/s$). These quantities provide insight 
into the sound attenuation, flow characteristics, fluid behavior, and the degree of conformal symmetry in the system. 

To compute the shear and bulk viscous coefficients for an expansion-induced anisotropic hot QCD medium at finite baryon asymmetry, we solve the relativistic Boltzmann transport equation in the relaxation time approximation within the kinetic theory framework. We also compare these viscous coefficients of baryon asymmetric matter with those of baryonless matter within the anisotropic environment in order to understand the influence of baryon asymmetry on the viscous properties of an anisotropic medium. In addition, this work involves the temperature-, chemical potential- and anisotropy parameter-dependent parton masses, distribution functions and dispersion relations within the quasiparticle model. 

This paper is structured as follows. Section 2 is devoted to the study of momentum transport in anisotropic hot QCD medium at finite baryon asymmetry, with particular emphasis on the shear and bulk viscosities. Section 3 involves the study of the corresponding dimensionless observables, such as the Prandtl number, the Reynolds number, specific shear viscosity and specific bulk viscosity. The quasiparticle model of partons in an anisotropic QCD medium with finite baryon asymmetry is discussed in section 4. Section 5 discusses the results obtained within this quasiparticle framework. Finally, in section 6, we present our conclusions. 

\section{Momentum transport phenomena for an anisotropic hot QCD medium at finite baryon asymmetry}
The rapid longitudinal expansion of the QGP during the early stages of ultrarelativistic heavy-ion collisions drives the system away from local momentum isotropy, resulting in a difference between the transverse and longitudinal momentum scales which can be characterized by an expansion-induced momentum anisotropy. In the regime of weak anisotropy ($\xi<1$) along a preferred direction $\mathbf{n}$, the distribution function in the anisotropic medium can be approximated by a deformed isotropic distribution, effectively characterized by a suppression of its high-momentum tail \cite{Romatschke:PRD68'2003}. Accordingly, the quark, antiquark and gluon distribution functions are modified through the following rescalings, 
\be\label{A.D.F.Q.}
&&f_f^\xi=\frac{N(\xi)}{e^{\beta\left(\sqrt{\omega_f^2+\xi(\mathbf{p}\cdot\mathbf{n})^2}-\mu_f\right)}+1} ~, \\ 
\label{A.D.F.A.}&&\bar{f}_f^\xi=\frac{N(\xi)}{e^{\beta\left(\sqrt{\omega_f^2+\xi(\mathbf{p}\cdot\mathbf{n})^2}+\mu_f\right)}+1} ~, \\ 
\label{A.D.F.G.}&&f_g^\xi=\frac{N(\xi)}{e^{\beta\sqrt{\omega_g^2+\xi(\mathbf{p}\cdot\mathbf{n})^2}}-1} 
~,\ee
respectively, where $N(\xi)=\sqrt{1+\xi}$ is the normalization factor \cite{Romatschke:PRD70'2004} and $\xi$ represents the anisotropy parameter. The Taylor series expansion up to $\mathcal{O}(\xi)$ gives the 
forms of $f_f^\xi$, $\bar{f}_f^\xi$ and $f_g^\xi$ as
\be\label{E.Q.}
&&f_f^\xi=f_f+\frac{\xi f_f}{2}-\frac{\xi\beta(\mathbf{p}\cdot\mathbf{n})^2}{2\omega_f}f_f\left(1-f_f\right), \\ 
\label{E.A.}&&\bar{f}_f^\xi=\bar{f}_f+\frac{\xi \bar{f}_f}{2}-\frac{\xi\beta(\mathbf{p}\cdot\mathbf{n})^2}{2\omega_f}\bar{f}_f\left(1-\bar{f}_f\right), \\ 
\label{E.G.}&&f_g^\xi=f_g+\frac{\xi f_g}{2}-\frac{\xi\beta(\mathbf{p}\cdot\mathbf{n})^2}{2\omega_g}f_g\left(1+f_g\right)
,\ee
where $f_f$, $\bar{f}_f$ and $f_g$ are the isotropic quark, antiquark and gluon distribution functions, which are expressed as
\be\label{I.D.F.Q.}
&&f_f=\frac{1}{e^{\beta\left(\omega_f-\mu_f\right)}+1} ~, \\ 
&&\label{I.D.F.A.Q.}\bar{f}_f=\frac{1}{e^{\beta\left(\omega_f+\mu_f\right)}+1} ~, \\ 
&&\label{I.D.F.G.}f_g=\frac{1}{e^{\beta\omega_g}-1}
~.\ee
Here, $\omega_f$ and $\omega_g$ are the energies of the quark (antiquark) of flavor $f$ and the gluon, respectively, in the baryon asymmetric hot QCD matter. The degree of anisotropy is characterized 
by the anisotropy parameter, which is defined in terms of the transverse and longitudinal momentum components as
\be\label{parameter}
\xi=\frac{\left\langle\mathbf{p}_T^2\right\rangle}{2\left\langle p_L^2\right\rangle}-1
~,\ee
where $p_L=\mathbf{p}\cdot\mathbf{n}$, $\mathbf{p}_T=\mathbf{p}-\mathbf{n}\cdot(\mathbf{p}\cdot\mathbf{n})$, $\mathbf{p}\equiv(\rm{p}\sin\theta\cos\phi,\rm{p}\sin\theta\sin\phi,\rm{p}\cos\theta)$, $\mathbf{n}=(\sin\alpha,0,\cos\alpha)$, $\alpha$ denotes the angle between the z-axis and the direction of anisotropy, $(\mathbf{p}\cdot\mathbf{n})^2=\rm{p}^2c(\alpha,\theta,\phi)=\rm{p}^2(\sin^2\alpha\sin^2\theta\cos^2\phi+\cos^2\alpha\cos^2\theta+\sin(2\alpha)\sin\theta\cos\theta\cos\phi)$. For $p_T\gg p_L$, the anisotropy parameter $\xi$ becomes positive, indicating a depletion of particles with large momentum components along the $\mathbf{n}$ direction due to the rapid longitudinal expansion compared with the transverse expansion of matter. 

The momentum transfer and viscous properties in the baryon asymmetric medium could be affected by the weak-momentum anisotropy arising due to the asymptotic expansion of matter. The influence of this anisotropy can 
be quantified by examining the shear and bulk viscosities of baryon asymmetric hot QCD matter in the anisotropic regime. When an anisotropic hot medium composed of quarks, antiquarks, and gluons with finite baryon asymmetry, initially in thermal equilibrium, is subjected to a small external perturbation, both the energy-momentum tensor and the parton distribution functions deviate infinitesimally from their equilibrium forms. Their infinitesimal deviations can be expressed as $T^{\mu\nu}\rightarrow {T^\prime}^{\mu\nu}=T^{\mu\nu}+\Delta T^{\mu\nu}$, $f^\xi_f\rightarrow f_f^{\prime \xi}=f^\xi_f+\delta f^\xi_f$, $\bar{f}_f^\xi\rightarrow \bar{f}_f^{\prime \xi}=\bar{f}_f^\xi+\delta \bar{f}_f^\xi$ and $f^\xi_g\rightarrow f_g^{\prime \xi}=f^\xi_g+\delta f^\xi_g$, where $\Delta T^{\mu\nu}$ represents the nonequilibrium part of the energy-momentum tensor, and $\delta f^\xi_f$, $\delta \bar{f}_f^\xi$ and $\delta f^\xi_g$ denote the infinitesimal changes in the quark, antiquark and gluon distribution functions, respectively. The energy-momentum tensor for a 
slightly nonequilibrium medium is written as
\be\label{N.E.M.}
{T^\prime}^{\mu\nu}=\int\frac{d^3{\rm p}}{(2\pi)^3}p^\mu p^\nu \left[\sum_f g_f\frac{\left(f_f^{\prime \xi}+\bar{f}_f^{\prime \xi}\right)}{{\omega_f}}+g_g\frac{f_g^{\prime \xi}}{\omega_g}\right]
,\ee
where the subscript $f$ denotes quark flavor. In eq. \eqref{N.E.M.}, $g_f$ and $g_g$ represent the degeneracy factors for quarks and gluons, respectively. Similarly, we can write $\Delta T^{\mu\nu}$ as
\be\label{em1}
\Delta T^{\mu\nu}=\int\frac{d^3{\rm p}}{(2\pi)^3}p^\mu p^\nu \left[\sum_f g_f\frac{\left(\delta f^\xi_f+\delta \bar{f}_f^\xi\right)}{{\omega_f}}+g_g\frac{\delta f^\xi_g}{\omega_g}\right]
.\ee
For the calculation of the shear and bulk viscosities, it is necessary to obtain the infinitesimal changes of the parton distribution functions appearing in eq. \eqref{em1}. To determine these infinitesimal changes, we solve the relativistic Boltzmann transport equation (RBTE) using the relaxation time approximation within the 
kinetic theory approach. 

In evaluating the shear and bulk viscous coefficients, the electromagnetic field contribution can be excluded from the relativistic Boltzmann transport equation. In the relaxation time approximation method, the relativistic Boltzmann transport equations for quarks, antiquarks and gluons are expressed as
\be\label{R.B.T.E.Q (1)}
p^\mu\partial_\mu f_f^{\prime \xi}=-\frac{p_\nu u^\nu}{\tau_f}\delta f_f^\xi, \\ 
\label{R.B.T.E.A (1)}p^\mu\partial_\mu \bar{f}_f^{\prime \xi}=-\frac{p_\nu u^\nu}{\tau\bar{f}}\delta \bar{f}_f^\xi, \\ 
\label{R.B.T.E.G (1)}p^\mu\partial_\mu f_g^{\prime \xi} = -\frac{p_\nu u^\nu}{\tau_g}\delta f_g^\xi
,\ee
respectively, where the relaxation times for quarks (antiquarks), $\tau_f$ ($\tau_{\bar{f}}$) and gluons, $\tau_g$ are momentum-independent with the following \cite{Hosoya:NPB250'1985} forms, 
\be\label{R.T.Q}
\tau_{f(\bar{f})} &=& \frac{1}{5.1T\alpha_s^2\log\left(1/\alpha_s\right)\left[1+0.12 
(2N_f+1)\right]} ~, \\ 
\label{R.T.G}\tau_g &=& \frac{1}{22.5T\alpha_s^2\log\left(1/\alpha_s\right)\left[1+0.06N_f\right]}
~.\ee
From equations \eqref{R.B.T.E.Q (1)}, \eqref{R.B.T.E.A (1)} and \eqref{R.B.T.E.G (1)}, the infinitesimal changes in the parton distribution functions are obtained as
\be
\delta f_f^\xi=-\frac{\tau_f p^\mu\partial_\mu f_f^{\prime \xi}}{p_\nu u^\nu}, \\ 
\delta \bar{f}_f^\xi=-\frac{\tau_{\bar{f}} p^\mu\partial_\mu \bar{f}_f^{\prime \xi}}{p_\nu u^\nu}, \\ 
\delta f_g^\xi=-\frac{\tau_g p^\mu\partial_\mu f_g^{\prime \xi}}{p_\nu u^\nu}
.\ee
Utilizing the values of $\delta f_f^\xi$, $\delta \bar{f}_f^\xi$ and $\delta f_g^\xi$ in eq. \eqref{em1}, we get 
\be\label{em2 (1)}
\Delta T^{\mu\nu}=-\int\frac{d^3{\rm p}}{(2\pi)^3}\frac{p^\mu p^\nu}{p_\nu u^\nu} \left[\sum_f g_f\left(\frac{\tau_f p^\mu\partial_\mu f_f^{\prime \xi}}{\omega_f}+\frac{\tau_{\bar{f}} p^\mu\partial_\mu \bar{f}_f^{\prime \xi}}{\omega_f}\right)+g_g\frac{\tau_g p^\mu\partial_\mu f_g^{\prime \xi}}{\omega_g}\right]
,\ee
where the partial derivative is decomposed as $\partial_\mu=u_\mu D+\nabla_\mu$, where $D=u^\mu\partial_\mu$. In the local rest frame, the distribution functions can be expanded in terms of the gradients of temperature and flow velocity. Accordingly, the partial derivatives of the nonequilibrium distribution functions for quarks, antiquarks and gluons are respectively obtained as
\begin{eqnarray}
\nonumber\partial_\mu f_f^{\prime \xi} &=& \beta f_f\left(1-f_f\right)\left\lbrace1+\frac{\xi}{2}-\frac{\xi\beta\rm{p}^2c(\alpha,\theta,\phi)}{2\omega_f}\left(1-2f_f\right)\right\rbrace
\left[u_\alpha p^\alpha u_\mu\frac{DT}{T}+u_\alpha p^\alpha\frac{\nabla_\mu T}{T}\right. \\ && \left.\nonumber -u_\mu p^\alpha Du_\alpha-p^\alpha\nabla_\mu u_\alpha+T\partial_\mu\left(\frac{\mu}{T}\right)\right]+\frac{\xi\beta\rm{p}^2c(\alpha,\theta,\phi)}{2\omega_f}f_f\left(1-f_f\right) \\ && \times\left[u_\mu\frac{DT}{T}+\frac{\nabla_\mu T}{T}\right]+\frac{\xi\beta\rm{p}^2c(\alpha,\theta,\phi)}{2\omega^2_f}f_f\left(1-f_f\right)\left[u_\mu p^\alpha Du_\alpha+p^\alpha\nabla_\mu u_\alpha\right]
,\end{eqnarray}
\begin{eqnarray}
\nonumber\partial_\mu \bar{f}_f^{\prime \xi} &=& \beta \bar{f}_f\left(1-\bar{f}_f\right)\left\lbrace1+\frac{\xi}{2}-\frac{\xi\beta\rm{p}^2c(\alpha,\theta,\phi)}{2\omega_f}\left(1-2\bar{f}_f\right)\right\rbrace
\left[u_\alpha p^\alpha u_\mu\frac{DT}{T}+u_\alpha p^\alpha\frac{\nabla_\mu T}{T}\right. \\ && \left.\nonumber -u_\mu p^\alpha Du_\alpha-p^\alpha\nabla_\mu u_\alpha-T\partial_\mu\left(\frac{\mu}{T}\right)\right]+\frac{\xi\beta\rm{p}^2c(\alpha,\theta,\phi)}{2\omega_f}\bar{f}_f\left(1-\bar{f}_f\right) \\ && \times\left[u_\mu\frac{DT}{T}+\frac{\nabla_\mu T}{T}\right]+\frac{\xi\beta\rm{p}^2c(\alpha,\theta,\phi)}{2\omega^2_f}\bar{f}_f\left(1-\bar{f}_f\right)\left[u_\mu p^\alpha Du_\alpha+p^\alpha\nabla_\mu u_\alpha\right]
,\end{eqnarray}
\begin{eqnarray}
\nonumber\partial_\mu f_g^{\prime \xi} &=& \beta f_g\left(1+f_g\right)\left\lbrace1+\frac{\xi}{2}-\frac{\xi\beta\rm{p}^2c(\alpha,\theta,\phi)}{2\omega_g}\left(1+2f_g\right)\right\rbrace
\left[u_\alpha p^\alpha u_\mu\frac{DT}{T}+u_\alpha p^\alpha\frac{\nabla_\mu T}{T}\right. \\ && \left.\nonumber -u_\mu p^\alpha Du_\alpha-p^\alpha\nabla_\mu u_\alpha\right]+\frac{\xi\beta\rm{p}^2c(\alpha,\theta,\phi)}{2\omega_g}f_g\left(1+f_g\right)\left[u_\mu\frac{DT}{T}+\frac{\nabla_\mu T}{T}\right] \\ && +\frac{\xi\beta\rm{p}^2c(\alpha,\theta,\phi)}{2\omega^2_g}f_g\left(1+f_g\right)\left[u_\mu p^\alpha Du_\alpha+p^\alpha\nabla_\mu u_\alpha\right]
.\end{eqnarray}
Substituting the expressions of $\partial_\mu f_f^{\prime \xi}$, $\partial_\mu \bar{f}_f^{\prime \xi}$ and $\partial_\mu f_g^{\prime \xi}$ in eq. \eqref{em2 (1)} and using $\frac{DT}{T}=-\left(\frac{\partial P}{\partial \varepsilon}\right)\nabla_\alpha u^\alpha$ and $Du_\alpha=\frac{\nabla_\alpha P}{\varepsilon+P}$ 
from the energy-momentum conservation, we get 
\be\label{em3 (1)}
\nonumber\Delta T^{\mu\nu} &=& \sum_f g_f\int\frac{d^3{\rm p}}{(2\pi)^3}\frac{p^\mu p^\nu\tau_f\beta}{\omega_f}f_f\left(1-f_f\right)\left[\left\lbrace1+\frac{\xi}{2}-\frac{\xi\beta\rm{p}^2c(\alpha,\theta,\phi)}{2\omega_f}\left(1-2f_f\right)\right\rbrace\right. \\ && \left.\nonumber\times\left\lbrace\omega_f\left(\frac{\partial P}{\partial \varepsilon}\right)\nabla_\alpha u^\alpha+p^\alpha\left(\frac{\nabla_\alpha P}{\varepsilon+P}-\frac{\nabla_\alpha T}{T}\right)-\frac{Tp^\alpha}{\omega_f}\partial_\alpha\left(\frac{\mu}{T}\right)+\frac{p^\alpha p^\beta}{\omega_f}\nabla_\alpha u_\beta\right\rbrace\right. \\ && \left.\nonumber+\frac{\xi\rm{p}^2c(\alpha,\theta,\phi)}{2\omega^2_f}\left\lbrace\omega_f\left(\frac{\partial P}{\partial \varepsilon}\right)\nabla_\alpha u^\alpha-p^\alpha\left(\frac{\nabla_\alpha P}{\varepsilon+P}+\frac{\nabla_\alpha T}{T}\right)-\frac{p^\alpha p^\beta}{\omega_f}\nabla_\alpha u_\beta\right\rbrace\right] \\ && \nonumber+\sum_f g_f\int\frac{d^3{\rm p}}{(2\pi)^3}\frac{p^\mu p^\nu\tau_{\bar{f}}\beta}{\omega_f}\bar{f}_f\left(1-\bar{f}_f\right)\left[\left\lbrace1+\frac{\xi}{2}-\frac{\xi\beta\rm{p}^2c(\alpha,\theta,\phi)}{2\omega_f}\left(1-2\bar{f}_f\right)\right\rbrace\right. \\ && \left.\nonumber\times\left\lbrace\omega_f\left(\frac{\partial P}{\partial \varepsilon}\right)\nabla_\alpha u^\alpha+p^\alpha\left(\frac{\nabla_\alpha P}{\varepsilon+P}-\frac{\nabla_\alpha T}{T}\right)+\frac{Tp^\alpha}{\omega_f}\partial_\alpha\left(\frac{\mu}{T}\right)+\frac{p^\alpha p^\beta}{\omega_f}\nabla_\alpha u_\beta\right\rbrace\right. \\ && \left.\nonumber+\frac{\xi\rm{p}^2c(\alpha,\theta,\phi)}{2\omega^2_f}\left\lbrace\omega_f\left(\frac{\partial P}{\partial \varepsilon}\right)\nabla_\alpha u^\alpha-p^\alpha\left(\frac{\nabla_\alpha P}{\varepsilon+P}+\frac{\nabla_\alpha T}{T}\right)-\frac{p^\alpha p^\beta}{\omega_f}\nabla_\alpha u_\beta\right\rbrace\right] \\ && \nonumber+g_g\int\frac{d^3{\rm p}}{(2\pi)^3}\frac{p^\mu p^\nu\tau_g\beta}{\omega_g}f_g\left(1+f_g\right)\left[\left\lbrace1+\frac{\xi}{2}-\frac{\xi\beta\rm{p}^2c(\alpha,\theta,\phi)}{2\omega_g}\left(1+2f_g\right)\right\rbrace\right. \\ && \left.\nonumber\times\left\lbrace\omega_g\left(\frac{\partial P}{\partial \varepsilon}\right)\nabla_\alpha u^\alpha+p^\alpha\left(\frac{\nabla_\alpha P}{\varepsilon+P}-\frac{\nabla_\alpha T}{T}\right)+\frac{p^\alpha p^\beta}{\omega_g}\nabla_\alpha u_\beta\right\rbrace\right. \\ && \left.+\frac{\xi\rm{p}^2c(\alpha,\theta,\phi)}{2\omega^2_g}\left\lbrace\omega_g\left(\frac{\partial P}{\partial \varepsilon}\right)\nabla_\alpha u^\alpha-p^\alpha\left(\frac{\nabla_\alpha P}{\varepsilon+P}+\frac{\nabla_\alpha T}{T}\right)-\frac{p^\alpha p^\beta}{\omega_g}\nabla_\alpha u_\beta\right\rbrace\right]
.\ee
It is possible to calculate the pressure and the energy density from their relations with the 
energy-momentum tensor as $P=-\Delta_{\mu\nu}T^{\mu\nu}/3$ and $\varepsilon=u_\mu T^{\mu\nu}u_\nu$, respectively, where $\Delta_{\mu\nu}=g_{\mu\nu}-u_\mu u_\nu$ denotes the projection tensor. Since $\Delta T^{00}=0$, only the spatial components of $\Delta T^{\mu\nu}$ depend on the velocity gradient in the local rest frame. From eq. \eqref{em3 (1)}, the spatial components of $\Delta T^{\mu\nu}$ are written as
\be\label{em4 (1)}
\nonumber\Delta T^{ij} &=& \sum_f g_f\int\frac{d^3{\rm p}}{(2\pi)^3}\frac{p^i p^j\tau_f\beta}{\omega_f}f_f\left(1-f_f\right)\left[\left\lbrace\left(1+\frac{\xi}{2}-\frac{\xi\beta\rm{p}^2c(\alpha,\theta,\phi)}{2\omega_f}\left(1-2f_f\right)\right)\right.\right. \\ && \left.\left.\nonumber\times\left(\omega_f\left(\frac{\partial P}{\partial \varepsilon}\right)-\frac{\rm p^2}{3\omega_f}\right)+\frac{\xi\rm{p}^2c(\alpha,\theta,\phi)}{2\omega^2_f}\left(\omega_f\left(\frac{\partial P}{\partial \varepsilon}\right)+\frac{\rm p^2}{3\omega_f}\right)\right\rbrace\partial_lu^l\right. \\ && \left.\nonumber -\left\lbrace1+\frac{\xi}{2}-\frac{\xi\beta\rm{p}^2c(\alpha,\theta,\phi)}{2\omega_f}\left(1-2f_f\right)-\frac{\xi\rm{p}^2c(\alpha,\theta,\phi)}{2\omega^2_f}\right\rbrace\frac{p^kp^l}{2\omega_f}W_{kl}\right. \\ && \left.\nonumber+\left\lbrace1+\frac{\xi}{2}-\frac{\xi\beta\rm{p}^2c(\alpha,\theta,\phi)}{2\omega_f}\left(1-2f_f\right)\right\rbrace\left\lbrace p^k\left(\frac{\partial_k P}{\varepsilon+P}-\frac{\partial_k T}{T} \right)-\frac{Tp^k}{\omega_f}\partial_k\left(\frac{\mu}{T}\right)\right\rbrace\right. \\ && \left.\nonumber -\frac{\xi\rm{p}^2c(\alpha,\theta,\phi)}{2\omega^2_f}p^k\left(\frac{\partial_k P}{\varepsilon+P}+\frac{\partial_k T}{T} \right)\right] \\ && \nonumber+\sum_f g_f\int\frac{d^3{\rm p}}{(2\pi)^3}\frac{p^i p^j\tau_{\bar{f}}\beta}{\omega_f}\bar{f}_f\left(1-\bar{f}_f\right)\left[\left\lbrace\left(1+\frac{\xi}{2}-\frac{\xi\beta\rm{p}^2c(\alpha,\theta,\phi)}{2\omega_f}\left(1-2\bar{f}_f\right)\right)\right.\right. \\ && \left.\left.\nonumber\times\left(\omega_f\left(\frac{\partial P}{\partial \varepsilon}\right)-\frac{\rm p^2}{3\omega_f}\right)+\frac{\xi\rm{p}^2c(\alpha,\theta,\phi)}{2\omega^2_f}\left(\omega_f\left(\frac{\partial P}{\partial \varepsilon}\right)+\frac{\rm p^2}{3\omega_f}\right)\right\rbrace\partial_lu^l\right. \\ && \left.\nonumber -\left\lbrace1+\frac{\xi}{2}-\frac{\xi\beta\rm{p}^2c(\alpha,\theta,\phi)}{2\omega_f}\left(1-2\bar{f}_f\right)-\frac{\xi\rm{p}^2c(\alpha,\theta,\phi)}{2\omega^2_f}\right\rbrace\frac{p^kp^l}{2\omega_f}W_{kl}\right. \\ && \left.\nonumber+\left\lbrace1+\frac{\xi}{2}-\frac{\xi\beta\rm{p}^2c(\alpha,\theta,\phi)}{2\omega_f}\left(1-2\bar{f}_f\right)\right\rbrace\left\lbrace p^k\left(\frac{\partial_k P}{\varepsilon+P}-\frac{\partial_k T}{T} \right)+\frac{Tp^k}{\omega_f}\partial_k\left(\frac{\mu}{T}\right)\right\rbrace\right. \\ && \left.\nonumber -\frac{\xi\rm{p}^2c(\alpha,\theta,\phi)}{2\omega^2_f}p^k\left(\frac{\partial_k P}{\varepsilon+P}+\frac{\partial_k T}{T} \right)\right] \\ && \nonumber+g_g\int\frac{d^3{\rm p}}{(2\pi)^3}\frac{p^i p^j\tau_g\beta}{\omega_g}f_g\left(1+f_g\right)\left[\left\lbrace\left(1+\frac{\xi}{2}-\frac{\xi\beta\rm{p}^2c(\alpha,\theta,\phi)}{2\omega_g}\left(1+2f_g\right)\right)\right.\right. \\ && \left.\left.\nonumber\times\left(\omega_g\left(\frac{\partial P}{\partial \varepsilon}\right)-\frac{\rm p^2}{3\omega_g}\right)+\frac{\xi\rm{p}^2c(\alpha,\theta,\phi)}{2\omega^2_g}\left(\omega_g\left(\frac{\partial P}{\partial \varepsilon}\right)+\frac{\rm p^2}{3\omega_g}\right)\right\rbrace\partial_lu^l\right. \\ && \left.\nonumber -\left\lbrace1+\frac{\xi}{2}-\frac{\xi\beta\rm{p}^2c(\alpha,\theta,\phi)}{2\omega_g}\left(1+2f_g\right)-\frac{\xi\rm{p}^2c(\alpha,\theta,\phi)}{2\omega^2_g}\right\rbrace\frac{p^kp^l}{2\omega_g}W_{kl}\right. \\ && \left.\nonumber+\left\lbrace1+\frac{\xi}{2}-\frac{\xi\beta\rm{p}^2c(\alpha,\theta,\phi)}{2\omega_g}\left(1+2f_g\right)\right\rbrace p^k\left(\frac{\partial_k P}{\varepsilon+P}-\frac{\partial_k T}{T} \right)\right. \\ && \left.-\frac{\xi\rm{p}^2c(\alpha,\theta,\phi)}{2\omega^2_g}p^k\left(\frac{\partial_k P}{\varepsilon+P}+\frac{\partial_k T}{T} \right)\right]
,\ee
where we have used $\partial_k u_l=-\frac{1}{2}W_{kl}-\frac{1}{3}\delta_{kl}\partial_j u^j$ and $W_{kl}=\partial_k u_l+\partial_l u_k-\frac{2}{3}\delta_{kl}\partial_j u^j$. The coefficients of the traceless 
and the trace parts of the dissipative contribution of the energy-momentum tensor define the shear and 
bulk viscosities, respectively. In the first order dissipative hydrodynamics, the spatial components of the infinitesimal shift of the energy-momentum tensor are given \cite{Lifshitz:BOOK'1981,Hosoya:NPB250'1985,Landau:BOOK'1987} by
\be\label{definition}
\Delta T^{ij}=-\eta W^{ij}-\zeta\delta^{ij}\partial_l u^l
.\ee
After comparing equations \eqref{em4 (1)} and \eqref{definition}, we get the shear viscosity for an expansion-induced anisotropic medium at finite baryon asymmetry as
\begin{eqnarray}\label{aniso.eta}
\nonumber\eta &=& \frac{\beta}{30\pi^2}\sum_f g_f \int d{\rm p}~\frac{{\rm p}^6}{\omega^2_f}\left[\tau_f f_f\left(1-f_f\right)+\tau_{\bar{f}}\bar{f_f}\left(1-\bar{f_f}\right)\right] \\ && \nonumber+\frac{\xi\beta}{60\pi^2}\sum_f g_f \int d{\rm p}~\frac{{\rm p}^6}{\omega^2_f}\left[\tau_f f_f\left(1-f_f\right)+\tau_{\bar{f}}\bar{f_f}\left(1-\bar{f_f}\right)\right] \\ && \nonumber -\frac{\xi\beta^2}{180\pi^2}\sum_f g_f \int d{\rm p}~\frac{{\rm p}^8}{\omega^3_f}\left[\tau_f f_f\left(1-f_f\right)\left(1-2f_f\right)+\tau_{\bar{f}}\bar{f_f}\left(1-\bar{f_f}\right)\left(1-2\bar{f_f}\right)\right] \\ && \nonumber -\frac{\xi\beta}{180\pi^2}\sum_f g_f \int d{\rm p}~\frac{{\rm p}^8}{\omega^4_f}\left[\tau_f f_f\left(1-f_f\right)+\tau_{\bar{f}}\bar{f_f}\left(1-\bar{f_f}\right)\right] \\ && \nonumber+\frac{\beta}{30\pi^2} g_g \int d{\rm p}~\frac{{\rm p}^6}{\omega^2_g} ~ \tau_g f_g\left(1+f_g\right)+\frac{\xi\beta}{60\pi^2} g_g \int d{\rm p}~\frac{{\rm p}^6}{\omega^2_g} ~ \tau_g f_g\left(1+f_g\right) \\ && \nonumber -\frac{\xi\beta^2}{180\pi^2} g_g \int d{\rm p}~\frac{{\rm p}^8}{\omega^3_g} ~ \tau_g f_g\left(1+f_g\right)\left(1+2f_g\right) \\ && -\frac{\xi\beta}{180\pi^2} g_g \int d{\rm p}~\frac{{\rm p}^8}{\omega^4_g} ~ \tau_g f_g\left(1+f_g\right)
~.\end{eqnarray}
The above equation can be written in terms of the $\xi$-independent ($\eta^i$) and $\xi$-dependent ($\eta^\xi$) parts as
\begin{eqnarray}\label{aniso.eta1}
\nonumber\eta &=& \eta^i+\eta^\xi \\ &=& \nonumber\frac{\beta}{30\pi^2}\sum_f g_f \int d{\rm p}~\frac{{\rm p}^6}{\omega^2_f}\left[\tau_f f_f\left(1-f_f\right)+\tau_{\bar{f}}\bar{f_f}\left(1-\bar{f_f}\right)\right]+\frac{\beta}{30\pi^2} g_g \int d{\rm p}~\frac{{\rm p}^6}{\omega^2_g} ~ \tau_g f_g\left(1+f_g\right) \\ && \nonumber+\frac{\xi\beta}{60\pi^2}\sum_f g_f \int d{\rm p}~\frac{{\rm p}^6}{\omega^2_f}\left[\tau_f f_f\left(1-f_f\right)\left\lbrace1-\frac{\beta{\rm p}^2}{3\omega_f}\left(1-2f_f\right)-\frac{{\rm p}^2}{3\omega^2_f}\right\rbrace\right. \\ && \left.\nonumber+\tau_{\bar{f}}\bar{f_f}\left(1-\bar{f_f}\right)\left\lbrace1-\frac{\beta{\rm p}^2}{3\omega_f}\left(1-2\bar{f_f}\right)-\frac{{\rm p}^2}{3\omega^2_f}\right\rbrace\right] \\ && +\frac{\xi\beta}{60\pi^2} g_g \int d{\rm p}~\frac{{\rm p}^6}{\omega^2_g} ~ \tau_g f_g\left(1+f_g\right)\left\lbrace1-\frac{\beta{\rm p}^2}{3\omega_g}\left(1+2f_g\right)-\frac{{\rm p}^2}{3\omega^2_g}\right\rbrace
~.\end{eqnarray}
In a similar way, the bulk viscosity is determined after comparing equations \eqref{em4 (1)} and \eqref{definition} as
\begin{eqnarray}\label{aniso.zeta (1)}
\nonumber\zeta &=& \frac{1}{3}\sum_f g_f \int\frac{d^3{\rm p}}{(2\pi)^3}~\frac{{\rm p}^2}{\omega_f}\left[f_f\left(1-f_f\right)A_f+\bar{f_f}\left(1-\bar{f_f}\right)\bar{A}_f\right] \\ && +\frac{1}{3}g_g \int\frac{d^3{\rm p}}{(2\pi)^3}~\frac{{\rm p}^2}{\omega_g}f_g\left(1+f_g\right)A_g
~,\end{eqnarray}
where the quantities $A_f$, $\bar{A}_f$ and $A_g$ are given by
\begin{eqnarray}
\nonumber A_f &=& \beta\tau_f\left[\left(1+\frac{\xi}{2}-\frac{\xi\beta\rm{p}^2c(\alpha,\theta,\phi)}{2\omega_f}\left(1-2f_f\right)\right)\left(\frac{\rm p^2}{3\omega_f}-\omega_f\left(\frac{\partial P}{\partial \varepsilon}\right)\right)\right. \\ && \left.-\frac{\xi\rm{p}^2c(\alpha,\theta,\phi)}{2\omega^2_f}\left(\frac{\rm p^2}{3\omega_f}+\omega_f\left(\frac{\partial P}{\partial \varepsilon}\right)\right)\right], \\ \nonumber \bar{A}_f &=& \beta \tau_{\bar{f}}\left[\left(1+\frac{\xi}{2}-\frac{\xi\beta\rm{p}^2c(\alpha,\theta,\phi)}{2\omega_f}\left(1-2\bar{f_f}\right)\right)\left(\frac{\rm p^2}{3\omega_f}-\omega_f\left(\frac{\partial P}{\partial \varepsilon}\right)\right)\right. \\ && \left.-\frac{\xi\rm{p}^2c(\alpha,\theta,\phi)}{2\omega^2_f}\left(\frac{\rm p^2}{3\omega_f}+\omega_f\left(\frac{\partial P}{\partial \varepsilon}\right)\right)\right], \\
\nonumber A_g &=& \beta \tau_g\left[\left(1+\frac{\xi}{2}-\frac{\xi\beta\rm{p}^2c(\alpha,\theta,\phi)}{2\omega_g}\left(1+2f_g\right)\right)\left(\frac{\rm p^2}{3\omega_g}-\omega_g\left(\frac{\partial P}{\partial \varepsilon}\right)\right)\right. \\ && \left.-\frac{\xi\rm{p}^2c(\alpha,\theta,\phi)}{2\omega^2_g}\left(\frac{\rm p^2}{3\omega_g}+\omega_g\left(\frac{\partial P}{\partial \varepsilon}\right)\right)\right]
.\end{eqnarray}
For the validity of the Landau-Lifshitz condition ($\Delta T^{00}=0$) in the local rest frame, the quantities $A_f$, $\bar{A}_f$ and $A_g$ are replaced by $A_f\rightarrow A_f^\prime=A_f-b_f\omega_f$, 
$\bar{A}_f\rightarrow \bar{A}_f^\prime=\bar{A}_f-\bar{b}_f\omega_f$ and $A_g\rightarrow A_g^\prime=A_g-b_g\omega_g$, where $b_f$, $\bar{b}_f$ and $b_g$ are arbitrary constants associated with the particle number and energy conservations. Using the ``00'' component of $\Delta T^{\mu\nu}$ (eq. \eqref{em3 (1)}), the Landau-Lifshitz conditions for $A_f$, $\bar{A}_f$ and $A_g$ are expressed as
\begin{eqnarray}
&&\sum_f g_f\int\frac{d^3{\rm p}}{(2\pi)^3} ~ \omega_f f_f\left(1-f_f\right)\left(A_f-b_f\omega_f\right)=0 \label{A_i (1)} ~,~ \\ 
&&\sum_f g_f\int\frac{d^3{\rm p}}{(2\pi)^3} ~ \omega_f \bar{f_f}\left(1-\bar{f_f}\right)\left(\bar{A}_f-\bar{b}_f\omega_f\right)=0 \label{A_i.1 (1)} ~,~ \\ 
&&g_g\int\frac{d^3{\rm p}}{(2\pi)^3} ~ \omega_g f_g\left(1+f_g\right)\left(A_g-b_g\omega_g\right)=0 \label{A_g (1)}
~,\end{eqnarray}
respectively. The quantities $b_f$, $\bar{b}_f$ and $b_g$ are obtained by solving equations \eqref{A_i (1)}, \eqref{A_i.1 (1)} and \eqref{A_g (1)}. Replacing $A_f\rightarrow A_f^\prime$, $\bar{A}_f\rightarrow \bar{A}_f^\prime$ and $A_g\rightarrow A_g^\prime$ in eq. \eqref{aniso.zeta (1)} and then simplifying, we get the bulk viscosity for an expansion-induced anisotropic medium at finite baryon asymmetry as
\begin{eqnarray}\label{aniso.zeta}
\nonumber\zeta &=& \frac{\beta}{18\pi^2}\sum_f g_f \int d{\rm p}~{\rm p}^2\left[\frac{{\rm p}^2}{\omega_f}-3\left(\frac{\partial P}{\partial \varepsilon}\right)\omega_f\right]^2\left[\tau_f f_f\left(1-f_f\right)+\tau_{\bar{f}}\bar{f_f}\left(1-\bar{f_f}\right)\right] \\ && \nonumber+\frac{\xi\beta}{36\pi^2}\sum_f g_f \int d{\rm p}~{\rm p}^2\left[\frac{{\rm p}^2}{\omega_f}-3\left(\frac{\partial P}{\partial \varepsilon}\right)\omega_f\right]^2\left[\tau_f f_f\left(1-f_f\right)+\tau_{\bar{f}}\bar{f_f}\left(1-\bar{f_f}\right)\right] \\ && \nonumber -\frac{\xi\beta^2}{108\pi^2}\sum_f g_f \int d{\rm p}~\frac{{\rm p}^4}{\omega_f}\left[\frac{{\rm p}^2}{\omega_f}-3\left(\frac{\partial P}{\partial \varepsilon}\right)\omega_f\right]^2\left[\tau_f f_f\left(1-f_f\right)\left(1-2f_f\right)\right. \\ && \left.\nonumber+\tau_{\bar{f}}\bar{f_f}\left(1-\bar{f_f}\right)\left(1-2\bar{f_f}\right)\right] \\ && \nonumber -\frac{\xi\beta}{108\pi^2}\sum_f g_f \int d{\rm p}~\frac{{\rm p}^4}{\omega^2_f}\left[\frac{{\rm p}^4}{\omega^2_f}-9\left(\frac{\partial P}{\partial \varepsilon}\right)^2\omega^2_f\right]\left[\tau_f f_f\left(1-f_f\right)+\tau_{\bar{f}}\bar{f_f}\left(1-\bar{f_f}\right)\right] \\ && \nonumber+\frac{\beta}{18\pi^2}g_g\int d{\rm p}~{\rm p}^2\left[\frac{{\rm p}^2}{\omega_g}-3\left(\frac{\partial P}{\partial \varepsilon}\right)\omega_g\right]^2 \tau_g f_g\left(1+f_g\right) \\ && \nonumber+\frac{\xi\beta}{36\pi^2}g_g\int d{\rm p}~{\rm p}^2\left[\frac{{\rm p}^2}{\omega_g}-3\left(\frac{\partial P}{\partial \varepsilon}\right)\omega_g\right]^2 \tau_g f_g\left(1+f_g\right) \\ && \nonumber -\frac{\xi\beta^2}{108\pi^2}g_g\int d{\rm p}~\frac{{\rm p}^4}{\omega_g}\left[\frac{{\rm p}^2}{\omega_g}-3\left(\frac{\partial P}{\partial \varepsilon}\right)\omega_g\right]^2 \tau_g f_g\left(1+f_g\right)\left(1+2f_g\right) \\ && -\frac{\xi\beta}{108\pi^2}g_g\int d{\rm p}~\frac{{\rm p}^4}{\omega^2_g}\left[\frac{{\rm p}^4}{\omega^2_g}-9\left(\frac{\partial P}{\partial \varepsilon}\right)^2\omega^2_g\right]\tau_g f_g\left(1+f_g\right)
~.\end{eqnarray}
The above equation can be written in terms of the $\xi$-independent ($\zeta^i$) and $\xi$-dependent ($\zeta^\xi$) parts as
\begin{eqnarray}\label{aniso.zeta1}
\nonumber\zeta &=& \zeta^i+\zeta^\xi \\ &=& \nonumber\frac{\beta}{18\pi^2}\sum_f g_f \int d{\rm p}~{\rm p}^2\left[\frac{{\rm p}^2}{\omega_f}-3\left(\frac{\partial P}{\partial \varepsilon}\right)\omega_f\right]^2\left[\tau_f f_f\left(1-f_f\right)+\tau_{\bar{f}}\bar{f_f}\left(1-\bar{f_f}\right)\right] \\ && \nonumber+\frac{\beta}{18\pi^2}g_g\int d{\rm p}~{\rm p}^2\left[\frac{{\rm p}^2}{\omega_g}-3\left(\frac{\partial P}{\partial \varepsilon}\right)\omega_g\right]^2 \tau_g f_g\left(1+f_g\right) \\ && \nonumber+\frac{\xi\beta}{36\pi^2}\sum_f g_f \int d{\rm p}~{\rm p}^2\left[\frac{{\rm p}^2}{\omega_f}-3\left(\frac{\partial P}{\partial \varepsilon}\right)\omega_f\right]^2\left[\tau_f f_f\left(1-f_f\right)\left\lbrace1-\frac{\beta{\rm p}^2}{3\omega_f}\left(1-2f_f\right)\right\rbrace\right. \\ && \left.\nonumber+\tau_{\bar{f}}\bar{f_f}\left(1-\bar{f_f}\right)\left\lbrace1-\frac{\beta{\rm p}^2}{3\omega_f}\left(1-2\bar{f_f}\right)\right\rbrace\right] \\ && \nonumber -\frac{\xi\beta}{108\pi^2}\sum_f g_f \int d{\rm p}~\frac{{\rm p}^4}{\omega^2_f}\left[\frac{{\rm p}^4}{\omega^2_f}-9\left(\frac{\partial P}{\partial \varepsilon}\right)^2\omega^2_f\right]\left[\tau_f f_f\left(1-f_f\right)+\tau_{\bar{f}}\bar{f_f}\left(1-\bar{f_f}\right)\right] \\ && \nonumber+\frac{\xi\beta}{36\pi^2}g_g\int d{\rm p}~{\rm p}^2\left[\frac{{\rm p}^2}{\omega_g}-3\left(\frac{\partial P}{\partial \varepsilon}\right)\omega_g\right]^2 \tau_g f_g\left(1+f_g\right)\left\lbrace1-\frac{\beta{\rm p}^2}{3\omega_g}\left(1+2f_g\right)\right\rbrace \\ && -\frac{\xi\beta}{108\pi^2}g_g\int d{\rm p}~\frac{{\rm p}^4}{\omega^2_g}\left[\frac{{\rm p}^4}{\omega^2_g}-9\left(\frac{\partial P}{\partial \varepsilon}\right)^2\omega^2_g\right]\tau_g f_g\left(1+f_g\right)
~.\end{eqnarray}

\section{Observables}
This section is dedicated to the study of observables related to the shear and bulk viscous coefficients in a baryon asymmetric hot QCD medium in the presence of expansion-induced anisotropy. Particularly, 
subsection 3.1 studies the Prandtl number, subsection 3.2 explores the Reynolds number and subsection 3.3 discusses the specific shear and specific bulk viscosities. 

\subsection{Prandtl number}
Momentum diffusion and thermal diffusion are related to each other through the Prandtl number as
\begin{equation}\label{Pl}
{\rm Pr}=\frac{\eta/\rho}{\kappa/C_p}
~,\end{equation}
where $\rho$ represents the mass density, $C_p$ is the specific heat at constant pressure and $\kappa$ 
denotes the thermal conductivity. The mass density can be calculated from its relation with the number densities of quarks, antiquarks and gluons as
\begin{equation}
\rho=\sum_f m_f\left(n_f+\bar{n}_f\right)+m_gn_g
~,\end{equation}
where $m_f$ and $m_g$ denote the thermal masses of charged particles (quarks or antiquarks) with flavor $f$ and gluons, respectively. In the presence of both anisotropy and baryon asymmetry, the mass density is 
obtained as
\begin{eqnarray}\label{M.D.}
\nonumber\rho &=& \frac{1}{2\pi^2}\sum_f m_f g_f\int d{\rm p}~{\rm p}^2\left(f_f+\bar{f_f}\right)+\frac{\xi}{4\pi^2}\sum_f m_f g_f\int d{\rm p}~{\rm p}^2\left(f_f+\bar{f_f}\right) \\ && \nonumber -\frac{\xi\beta}{12\pi^2}\sum_f m_f g_f\int d{\rm p}~\frac{{\rm p}^4}{\omega_f}\left[f_f\left(1-f_f\right)+\bar{f_f}\left(1-\bar{f_f}\right)\right] \\ && \nonumber+\frac{1}{2\pi^2}m_g g_g\int d{\rm p}~{\rm p}^2f_g+\frac{\xi}{4\pi^2}m_g g_g\int d{\rm p}~{\rm p}^2f_g \\ && -\frac{\xi\beta}{12\pi^2}m_g g_g\int d{\rm p}~\frac{{\rm p}^4}{\omega_g}f_g\left(1+f_g\right)
.\end{eqnarray}
The specific heat at constant pressure can be determined from the energy-momentum tensor as
\be
C_P &=& \frac{\partial\left[u_\mu T^{\mu\nu}u_\nu-\left(g_{\mu\nu}-u_\mu u_\nu\right)T^{\mu\nu}/3\right]}{\partial T}
.\ee
At finite anisotropy and baryon asymmetry, the specific heat at constant pressure is calculated to be 
\be\label{S.P.}
\nonumber C_P &=& \frac{\beta^2}{2\pi^2}\sum_fg_f\int d{\rm p} ~ {\rm p}^2\left(\omega_f+\frac{{\rm p}^2}{3\omega_f}\right)\left[\left(\omega_f-\mu\right)f_f\left(1-f_f\right)+\left(\omega_f+\mu\right)\bar{f_f}\left(1-\bar{f_f}\right)\right] \\ && \nonumber+\frac{\xi\beta^2}{4\pi^2}\sum_fg_f\int d{\rm p} ~ {\rm p}^2\left(\omega_f+\frac{{\rm p}^2}{3\omega_f}\right)\left[\left(\omega_f-\mu\right)f_f\left(1-f_f\right)+\left(\omega_f+\mu\right)\bar{f_f}\left(1-\bar{f_f}\right)\right] \\ && \nonumber+\frac{\xi\beta^2}{12\pi^2}\sum_fg_f\int d{\rm p} ~ \frac{{\rm p}^4}{\omega_f}\left(\omega_f+\frac{{\rm p}^2}{3\omega_f}\right)\left[f_f\left(1-f_f\right)+\bar{f_f}\left(1-\bar{f_f}\right)\right] \\ && \nonumber -\frac{\xi\beta^3}{12\pi^2}\sum_fg_f\int d{\rm p} ~ \frac{{\rm p}^4}{\omega_f}\left(\omega_f+\frac{{\rm p}^2}{3\omega_f}\right)\left[\left(\omega_f-\mu\right)f_f\left(1-f_f\right)\left(1-2f_f\right)\right. \\ && \left.\nonumber+\left(\omega_f+\mu\right)\bar{f_f}\left(1-\bar{f_f}\right)\left(1-2\bar{f_f}\right)\right] \\ && \nonumber+\frac{\beta^2}{2\pi^2}g_g\int d{\rm p} ~ {\rm p}^2\left(\omega_g+\frac{{\rm p}^2}{3\omega_g}\right)\omega_gf_g\left(1+f_g\right) \\ && \nonumber+\frac{\xi\beta^2}{4\pi^2}g_g\int d{\rm p} ~ {\rm p}^2\left(\omega_g+\frac{{\rm p}^2}{3\omega_g}\right)\omega_gf_g\left(1+f_g\right) \\ && \nonumber+\frac{\xi\beta^2}{12\pi^2}g_g\int d{\rm p} ~ \frac{{\rm p}^4}{\omega_g}\left(\omega_g+\frac{{\rm p}^2}{3\omega_g}\right)f_g\left(1+f_g\right) \\ && -\frac{\xi\beta^3}{12\pi^2}g_g\int d{\rm p} ~ {\rm p}^4\left(\omega_g+\frac{{\rm p}^2}{3\omega_g}\right)f_g\left(1+f_g\right)\left(1+2f_g\right)
.\ee
The Prandtl number provides key insight into the relative roles of momentum diffusion and thermal diffusion in governing the sound attenuation within a medium. When Pr$<$1, thermal diffusion dominates over momentum diffusion, whereas the opposite behavior occurs for Pr$>$1. In this work, we 
evaluate the Prandtl number for a baryon asymmetric hot QCD medium in the presence of expansion-induced anisotropy. This analysis uses the thermal conductivity calculated in a similar environment of finite anisotropy and finite baryon asymmetry \cite{Rath:EPJA62'2026}. 

\subsection{Reynolds number}
The flow characteristics of the matter can be characterized by the Reynolds number through the following relation involving the kinematic viscosity (${\eta}/{\rho}$) as
\begin{equation}\label{Rl}
{\rm Re}=\frac{Lv}{\eta/\rho}
~,\end{equation}
where $L$, $v$ and $\rho$ denote the characteristic length of the system, the velocity of the flow and the mass density, respectively. By setting $L=4$ fm and $v\simeq 1$, and using the values of $\eta$ and $\rho$, we have determined the Reynolds number for a baryon asymmetric hot QCD medium in the presence of expansion-induced anisotropy. The value of the Reynolds number provides insight into the fluidity of the system. When it significantly exceeds unity, the nature of the flow is turbulent and when it is small, the medium behaves as a viscous system, exhibiting laminar flow characteristics. Similar to other external conditions, the expansion-induced anisotropy can conspicuously modify the Reynolds number, potentially altering the flow characteristics within this regime. The present work explores how expansion-induced anisotropy affects the Reynolds number 
of a baryon asymmetric hot QCD medium. 

\subsection{Specific shear viscosity and specific bulk viscosity}
In a medium, the specific shear ($\eta/s$) and specific bulk ($\zeta/s$) viscosities are important for characterizing the perfect fluid behavior and the degree of conformal symmetry. In an 
expansion-induced anisotropic medium, these ratios are governed primarily by how shear viscosity, bulk viscosity and entropy density respond to finite temperature, anisotropy and baryon asymmetry. The entropy density ($s$) can be obtained from the thermodynamic relation involving the energy-momentum tensor and the baryon density ($n_B$) as
\begin{eqnarray}\label{E.D.}
s=\beta\left[u_\mu T^{\mu\nu}u_\nu-\sum_{f}\mu_f n_B-\Delta_{\mu\nu}T^{\mu\nu}/3\right]
~.\end{eqnarray}
The baryon density for an expansion-induced anisotropic medium at finite baryon asymmetry is calculated to be 
\begin{eqnarray}
\nonumber n_B &=& \frac{1}{2\pi^2}\sum_f g_f\int d{\rm p}~{\rm p}^2\left(f_f-\bar{f_f}\right)+\frac{\xi}{4\pi^2}\sum_f g_f\int d{\rm p}~{\rm p}^2\left(f_f-\bar{f_f}\right) \\ && -\frac{\xi\beta}{12\pi^2}\sum_f g_f\int d{\rm p}~\frac{{\rm p}^4}{\omega_f}\left[f_f\left(1-f_f\right)-\bar{f_f}\left(1-\bar{f_f}\right)\right]
.\end{eqnarray}
The impact of the expansion-induced anisotropy on the specific shear and specific bulk viscosities of a baryon asymmetric hot QCD medium depends on how the shear viscosity, bulk viscosity and entropy density behave in a similar regime. 

\section{Quasiparticle model of an expansion-induced anisotropic hot QCD medium at finite baryon asymmetry}
The quasiparticle approach incorporates the effects of interactions in the QGP by assigning effective medium-dependent masses to its constituent partons. These masses reflect the modification of partonic propagation induced by the surrounding thermal medium, and depend on its thermodynamic state and momentum-space anisotropy. Under extreme conditions including anisotropies which encountered in the early stages of heavy-ion 
collisions, the resulting thermal masses can therefore differ significantly from their isotropic counterparts. Within the semiclassical transport framework, the squared thermal masses of quarks and gluons in baryon asymmetric hot QCD medium are expressed \cite{Kelly:PRD50'1994,Litim:PR364'2002} as
\be\label{Q.P.M.Q.(definition of quark mass)}
\nonumber m_{fT}^2 &=& \frac{g^2\left(N_c^2-1\right)}{2N_c}\int\frac{d^3{\rm p}}{(2\pi)^3} ~ \frac{1}{\rm p}\left[f_g+\frac{1}{2}\left(f_f+\bar{f}_f\right)\right] \\ &=& \frac{g^2\left(N_c^2-1\right)}{8\pi^2N_c}\int d{\rm p} ~ {\rm p}\left[\frac{\rm 2}{\left(e^{\beta\omega_g}-1\right)}+\frac{1}{\left(e^{\beta(\omega_f-\mu_f)}+1\right)}+\frac{1}{\left(e^{\beta(\omega_f+\mu_f)}+1\right)}\right], \\ 
\label{Q.P.M.G.(definition of gluon mass)}\nonumber m_{gT}^2 &=& \frac{g^2}{2}\left[-2N_c\int\frac{d^3{\rm p}}{(2\pi)^3} ~ \frac{\partial f_g}{\partial{\rm p}}-N_f\int\frac{d^3{\rm p}}{(2\pi)^3} ~ \left(\frac{\partial f_f}{\partial{\rm p}}+\frac{\partial \bar{f}_f}{\partial{\rm p}}\right)\right] \\ &=& \nonumber\frac{g^2N_c}{2\pi^2T}\int d{\rm p} ~ \frac{{\rm p}^3}{\omega_g}\frac{e^{\beta\omega_g}}{\left(e^{\beta\omega_g}-1\right)^2}+\frac{g^2N_f}{4\pi^2T}\int d{\rm p}~\frac{{\rm p}^3}{\omega_f}\left[\frac{e^{\beta(\omega_f-\mu_f)}}{\left(e^{\beta(\omega_f-\mu_f)}+1\right)^2}\right. \\ && \left.+\frac{e^{\beta(\omega_f+\mu_f)}}{\left(e^{\beta(\omega_f+\mu_f)}+1\right)^2}\right]
,\ee
respectively. Using the hard thermal loop approximation, the above equations can be evaluated. The simplified forms of quasiparticle masses (squared) of quark and gluon up to one-loop are obtained \cite{Braaten:PRD45'1992,Peshier:PRD66'2002,Blaizot:PRD72'2005} as
\be\label{Q.P.M.(Quark mass)}
&& m_{fT}^2=\frac{g^2T^2}{6}\left(1+\frac{\mu_f^2}{\pi^2T^2}\right), \\
&&\label{Q.P.M.(Gluon mass)}m_{gT}^2=\frac{g^2T^2}{6}\left(N_c+\frac{N_f}{2}+\frac{3}{2\pi^2T^2}\sum_f\mu_f^2\right)
,\ee
respectively. These quasiparticle masses are functions of both temperature and chemical potential. In this work, we use the $u$, $d$ and $s$ quark flavors, which are assigned equal chemical potentials, {\em i.e.} 
$\mu_f=\mu$. In the above equations, $g$ represents the one-loop running coupling, which is given \cite{Kapusta:BOOK'2006} by
\begin{eqnarray}
g^2=4\pi\alpha_s=\frac{48\pi^2}{\left(11N_c-2N_f\right)\ln\left({\Lambda^2}/{\Lambda_{\rm\overline{MS}}^2}\right)}
~.\end{eqnarray}
Here, $\Lambda_{\rm\overline{MS}}=0.176$ GeV, $\Lambda=2\pi\sqrt{T^2+\mu_f^2/\pi^2}$ for electrically charged particles (quarks and antiquarks) and $\Lambda=2 \pi T$ for gluons. The momentum-space anisotropy induced by the expansion modifies the quasiparticle masses of the constituent partons and, consequently, their dispersion relations. Therefore, the quasiparticle masses of quarks and gluons acquire anisotropy-dependent corrections. Incorporating these modifications, the expressions in equations \eqref{Q.P.M.Q.(definition of quark mass)} and \eqref{Q.P.M.G.(definition of gluon mass)} are respectively generalized to 
\be\label{Q.P.M.Q.(Anisotropy)}
m_{fT\xi}^2 &=& \frac{g^2\left(N_c^2-1\right)}{2N_c}\int\frac{d^3{\rm p}}{(2\pi)^3} ~ \frac{1}{\rm p}\left[f_g^\xi+\frac{1}{2}\left(f_f^\xi+\bar{f}_f^\xi\right)\right], \\ 
\label{Q.P.M.G.(Anisotropy)}
m_{gT\xi}^2 &=& \frac{g^2}{2}\left[-2N_c\int\frac{d^3{\rm p}}{(2\pi)^3} ~ \frac{\partial f_g^\xi}{\partial{\rm p}}-N_f\int\frac{d^3{\rm p}}{(2\pi)^3} ~ \left(\frac{\partial f_f^\xi}{\partial{\rm p}}+\frac{\partial \bar{f}_f^\xi}{\partial{\rm p}}\right)\right]
.\ee
Using the values of anisotropic distribution functions $f_f^\xi$, $\bar{f}_f^\xi$ and $f_g^\xi$ into the corresponding mass expressions and then carrying out the momentum integrations yields the squared 
quasiparticle masses of quark and gluon for an expansion-induced anisotropic hot QCD medium at finite baryon asymmetry as
\be\label{Q.P.M.Q.(Anisotropy 1)}
\nonumber m_{fT\xi}^2 &=& \frac{2g^2}{3\pi^2}\int d{\rm p} ~ {\rm p} \left[f_g+\frac{1}{2}\left(f_f+\bar{f}_f\right)\right]+\frac{\xi g^2}{3\pi^2}\int d{\rm p} ~ {\rm p} \left[f_g+\frac{1}{2}\left(f_f+\bar{f}_f\right)\right] \\ && -\frac{\xi\beta g^2}{9\pi^2}\int d{\rm p} ~ {\rm p^2} ~ f_g\left(1+f_g\right)-\frac{\xi\beta g^2}{18\pi^2}\int d{\rm p} ~ \frac{\rm p^3}{\omega_f}\left[f_f\left(1-f_f\right)+\bar{f}_f\left(1-\bar{f}_f\right)\right], \\ 
\label{Q.P.M.G.(Anisotropy 1)}
\nonumber m_{gT\xi}^2 &=& \frac{3\beta g^2}{2\pi^2}\left[\int d{\rm p} ~ {\rm p^2} ~ f_g\left(1+f_g\right)+\sum_f\int d{\rm p} ~ \frac{\rm p^3}{6\omega_f}\left\lbrace f_f\left(1-f_f\right)+\bar{f}_f\left(1-\bar{f}_f\right)\right\rbrace\right] \\ && \nonumber+\frac{3\xi\beta g^2}{4\pi^2}\int d{\rm p} ~ {\rm p^2} ~ f_g\left(1+f_g\right)+\frac{\xi\beta g^2}{4\pi^2}\int d{\rm p} ~ {\rm p^2} ~ f_g\left(1+f_g\right) \\ && \nonumber -\frac{\xi\beta^2 g^2}{4\pi^2}\int d{\rm p} ~ {\rm p^3}\left(1+2f_g\right)f_g\left(1+f_g\right) \\ && +\nonumber\sum_f\frac{\xi\beta g^2}{8\pi^2}\int d{\rm p} ~ \frac{\rm p^3}{\omega_f}\left[f_f\left(1-f_f\right)+\bar{f}_f\left(1-\bar{f}_f\right)\right] \\ && \nonumber+\sum_f\frac{\xi\beta g^2}{12\pi^2}\int d{\rm p} ~ \frac{\rm p^3}{\omega_f}\left[f_f\left(1-f_f\right)+\bar{f}_f\left(1-\bar{f}_f\right)\right] \\ && - \nonumber\sum_f\frac{\xi\beta g^2}{24\pi^2}\int d{\rm p} ~ \frac{\rm p^5}{\omega_f^3}\left[f_f\left(1-f_f\right)+\bar{f}_f\left(1-\bar{f}_f\right)\right] \\ && -\sum_f\frac{\xi\beta^2 g^2}{24\pi^2}\int d{\rm p} ~ \frac{\rm p^5}{\omega_f^2}\left[\left(1-2f_f\right)f_f\left(1-f_f\right)+\left(1-2\bar{f}_f\right)\bar{f}_f\left(1-\bar{f}_f\right)\right]
,\ee
respectively. The first terms on the right-hand sides of equations \eqref{Q.P.M.Q.(Anisotropy 1)} and \eqref{Q.P.M.G.(Anisotropy 1)} represent the quasiparticle masses (squared) of quark and gluon, 
respectively for an isotropic hot QCD medium at finite baryon asymmetry. Thus, we have 
\be\label{Q.P.M.Q.(Anisotropic medium)}
\nonumber m_{fT\xi}^2 &=& m_{fT}^2+\frac{\xi g^2}{3\pi^2}\int d{\rm p} ~ {\rm p} \left[f_g+\frac{1}{2}\left(f_f+\bar{f}_f\right)\right]-\frac{\xi\beta g^2}{9\pi^2}\int d{\rm p} ~ {\rm p^2} ~ f_g\left(1+f_g\right) \\ && -\frac{\xi\beta g^2}{18\pi^2}\int d{\rm p} ~ \frac{\rm p^3}{\omega_f}\left[f_f\left(1-f_f\right)+\bar{f}_f\left(1-\bar{f}_f\right)\right], \\ 
\label{Q.P.M.G.(Anisotropic medium)}
\nonumber m_{gT\xi}^2 &=& m_{gT}^2+\frac{3\xi\beta g^2}{4\pi^2}\int d{\rm p} ~ {\rm p^2} ~ f_g\left(1+f_g\right)+\frac{\xi\beta g^2}{4\pi^2}\int d{\rm p} ~ {\rm p^2} ~ f_g\left(1+f_g\right) \\ && \nonumber -\frac{\xi\beta^2 g^2}{4\pi^2}\int d{\rm p} ~ {\rm p^3}\left(1+2f_g\right)f_g\left(1+f_g\right) \\ && +\nonumber\sum_f\frac{\xi\beta g^2}{8\pi^2}\int d{\rm p} ~ \frac{\rm p^3}{\omega_f}\left[f_f\left(1-f_f\right)+\bar{f}_f\left(1-\bar{f}_f\right)\right] \\ && \nonumber+\sum_f\frac{\xi\beta g^2}{12\pi^2}\int d{\rm p} ~ \frac{\rm p^3}{\omega_f}\left[f_f\left(1-f_f\right)+\bar{f}_f\left(1-\bar{f}_f\right)\right] \\ && - \nonumber\sum_f\frac{\xi\beta g^2}{24\pi^2}\int d{\rm p} ~ \frac{\rm p^5}{\omega_f^3}\left[f_f\left(1-f_f\right)+\bar{f}_f\left(1-\bar{f}_f\right)\right] \\ && -\sum_f\frac{\xi\beta^2 g^2}{24\pi^2}\int d{\rm p} ~ \frac{\rm p^5}{\omega_f^2}\left[\left(1-2f_f\right)f_f\left(1-f_f\right)+\left(1-2\bar{f}_f\right)\bar{f}_f\left(1-\bar{f}_f\right)\right]
.\ee
These expressions demonstrate that, in an expansion-induced anisotropic QCD medium with finite baryon asymmetry, the quasiparticle masses acquire an explicit dependence on the anisotropy parameter in 
addition to their dependence on the temperature and the chemical potential. This anisotropy dependence consequently modifies the partonic dispersion relations, reflecting the altered propagation of quarks and gluons in the medium. 

\begin{figure}[]
\begin{center}
\begin{tabular}{c c}
\includegraphics[width=7.4cm]{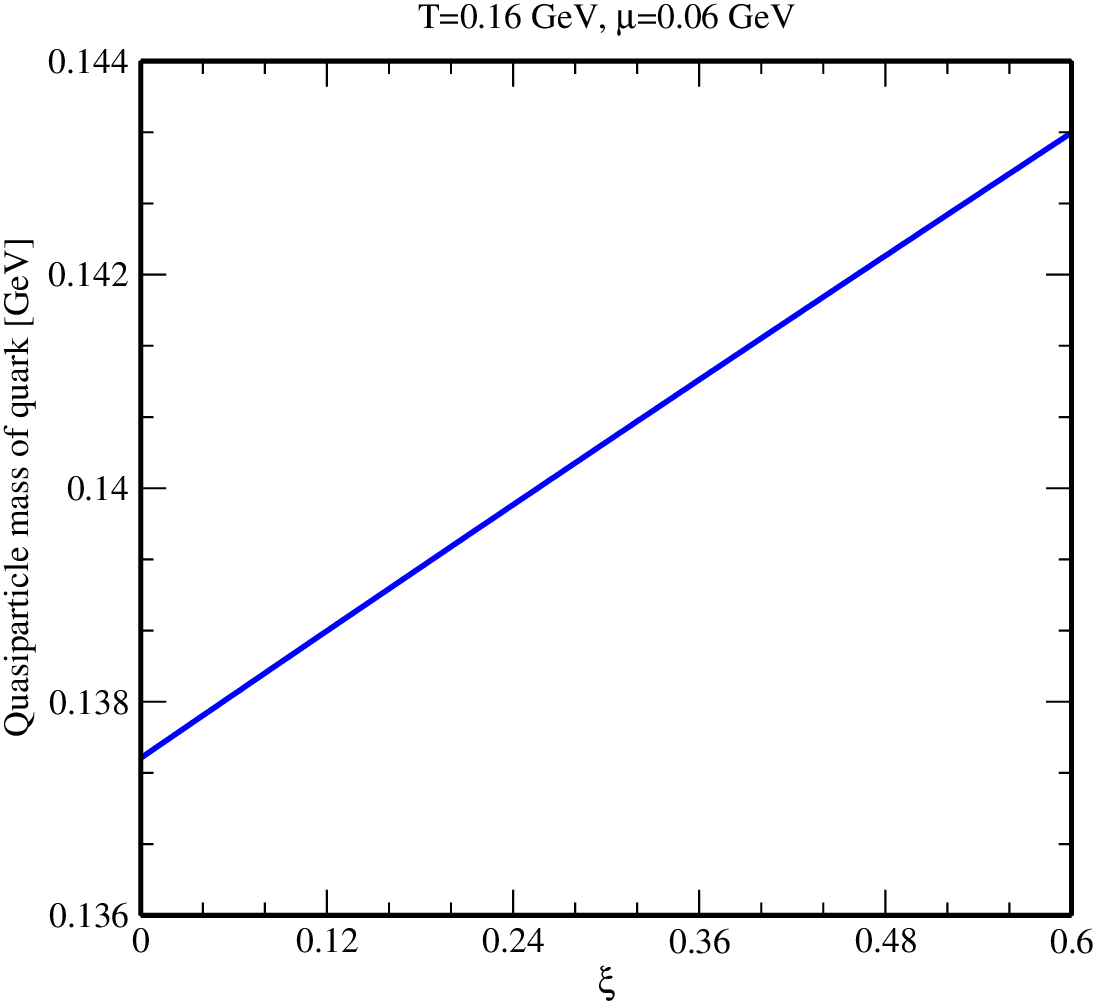}&
\hspace{0.73 cm}
\includegraphics[width=7.4cm]{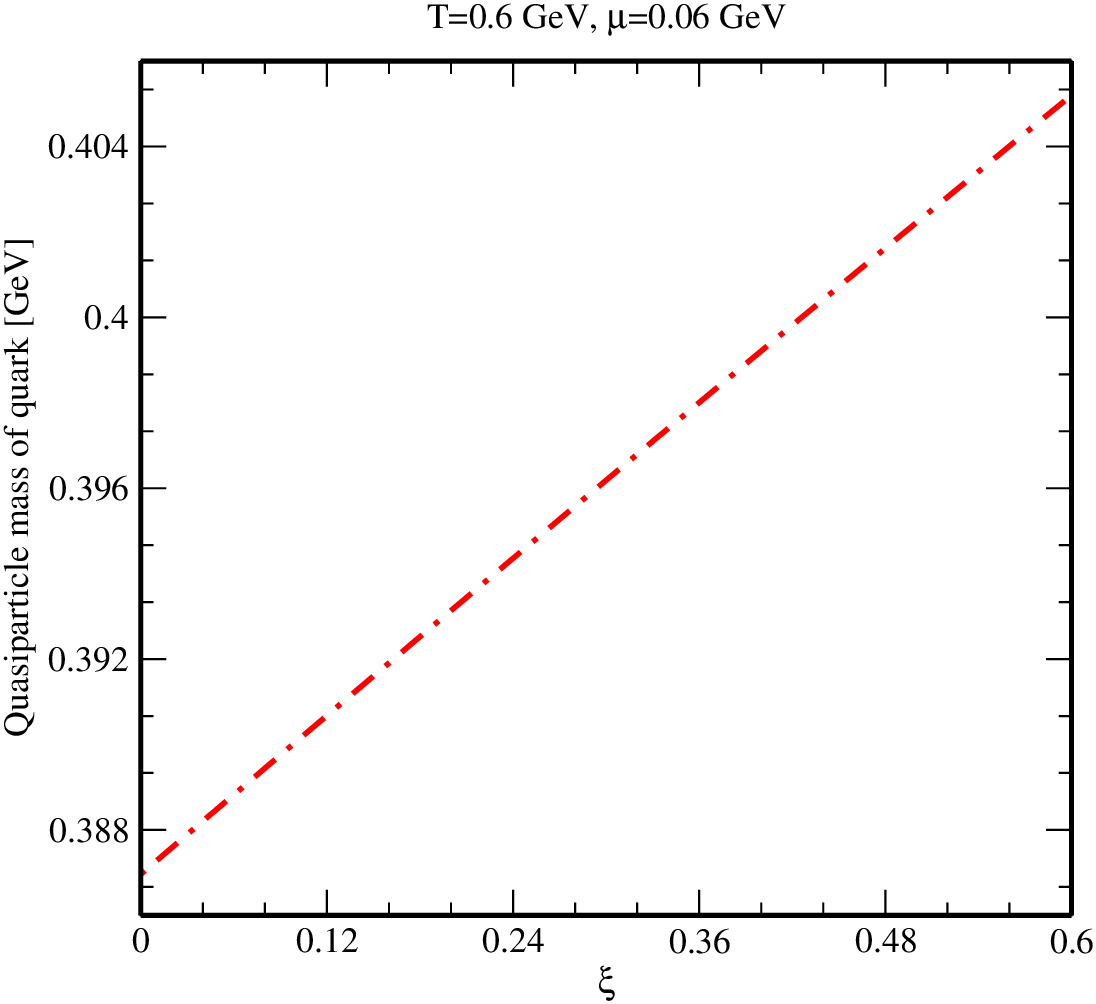} \\
a & b
\end{tabular}
\caption{Variations of the quasiparticle mass of quark with anisotropy parameter (a) at $T=0.16$ GeV and (b) at $T=0.6$ GeV for a fixed chemical potential.}\label{Fig.mq}
\end{center}
\end{figure}

\begin{figure}[]
\begin{center}
\begin{tabular}{c c}
\includegraphics[width=7.4cm]{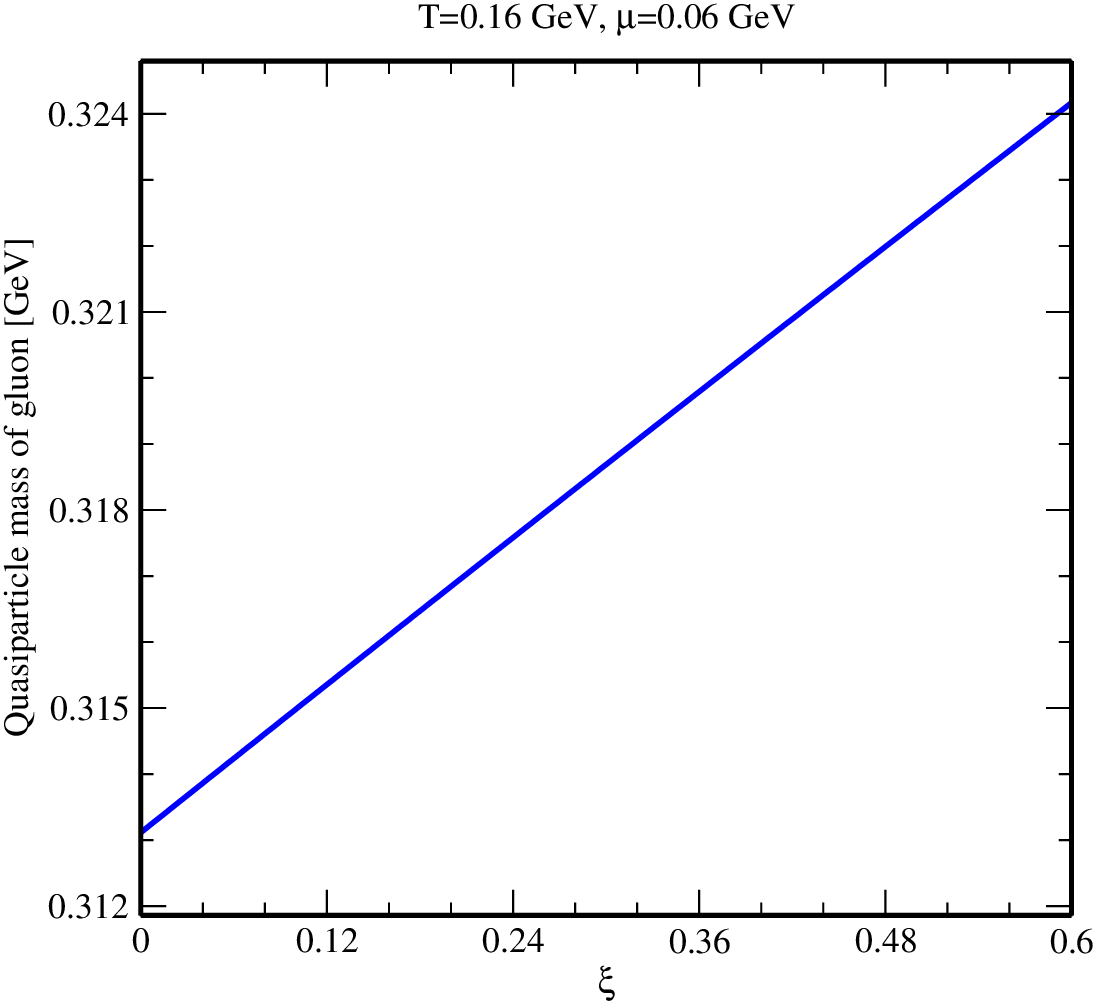}&
\hspace{0.73 cm}
\includegraphics[width=7.4cm]{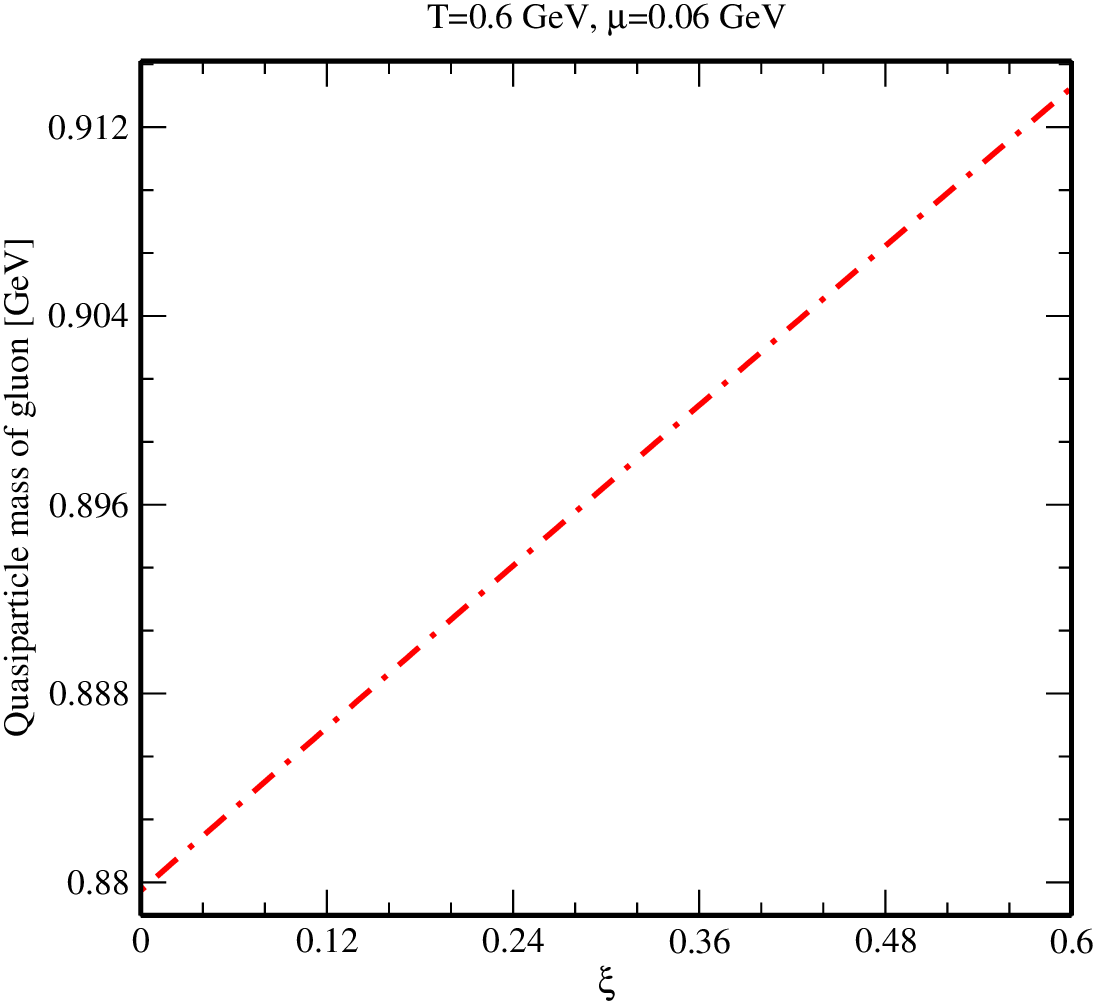} \\
a & b
\end{tabular}
\caption{Variations of the quasiparticle mass of gluon with anisotropy parameter (a) at $T=0.16$ GeV and (b) at $T=0.6$ GeV for a fixed chemical potential.}\label{Fig.mg}
\end{center}
\end{figure}

To understand the shear and bulk viscosities and their associated observables in an expansion-induced anisotropic medium, it is necessary to establish how the underlying partonic distributions are 
modified by the momentum anisotropy. In kinetic theory, the transport coefficients are determined by the microscopic properties of the medium, which enter through the parton distribution functions and their dispersion relations. The effect of the expansion-induced anisotropy on these quantities is therefore examined by studying the anisotropic dependence of the distribution functions at different temperatures. In the quasiparticle description, this anisotropic dependence is incorporated through the modified effective masses 
of the partons, which in turn alter their dispersion relations and consequently the transport properties of the medium. 

\begin{figure}[]
\begin{center}
\begin{tabular}{c c}
\includegraphics[width=7.4cm]{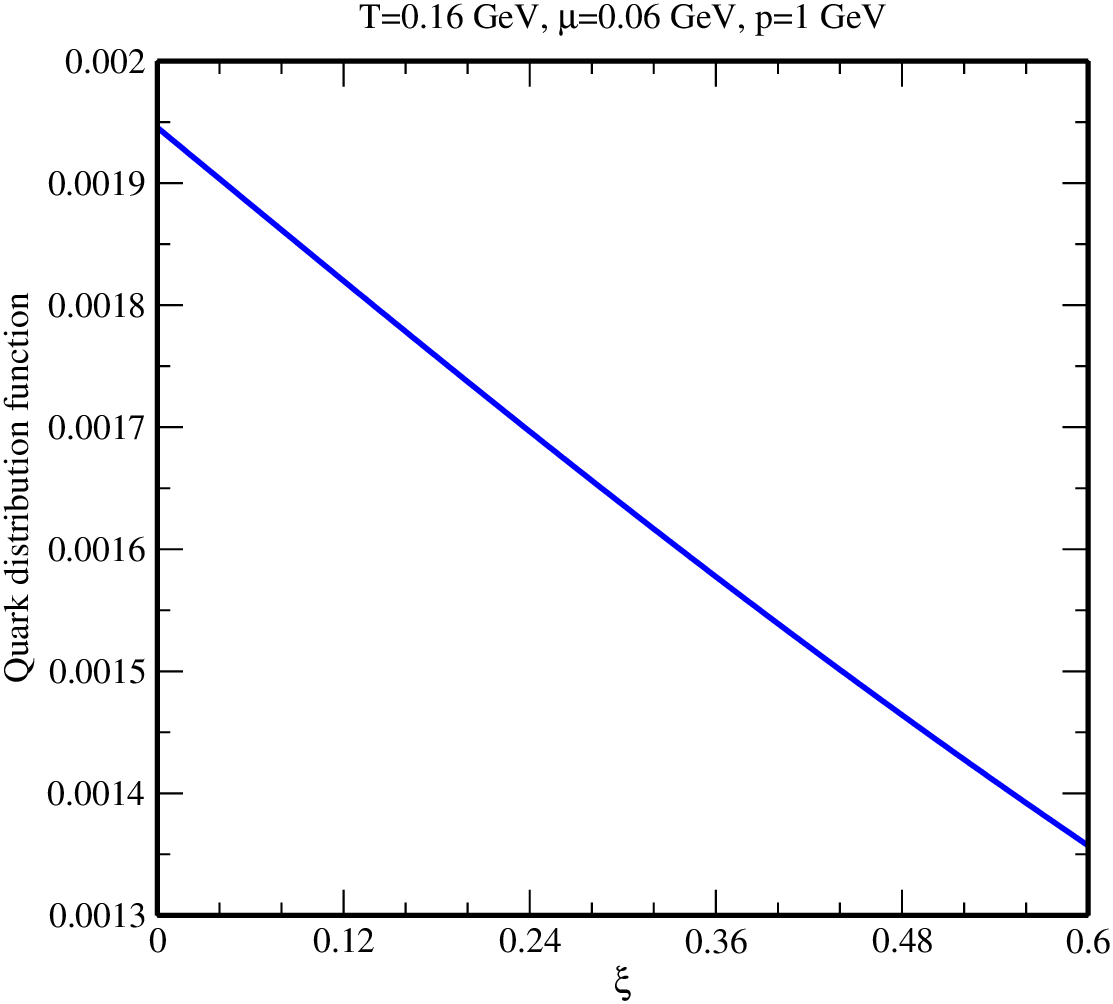}&
\hspace{0.73 cm}
\includegraphics[width=7.4cm]{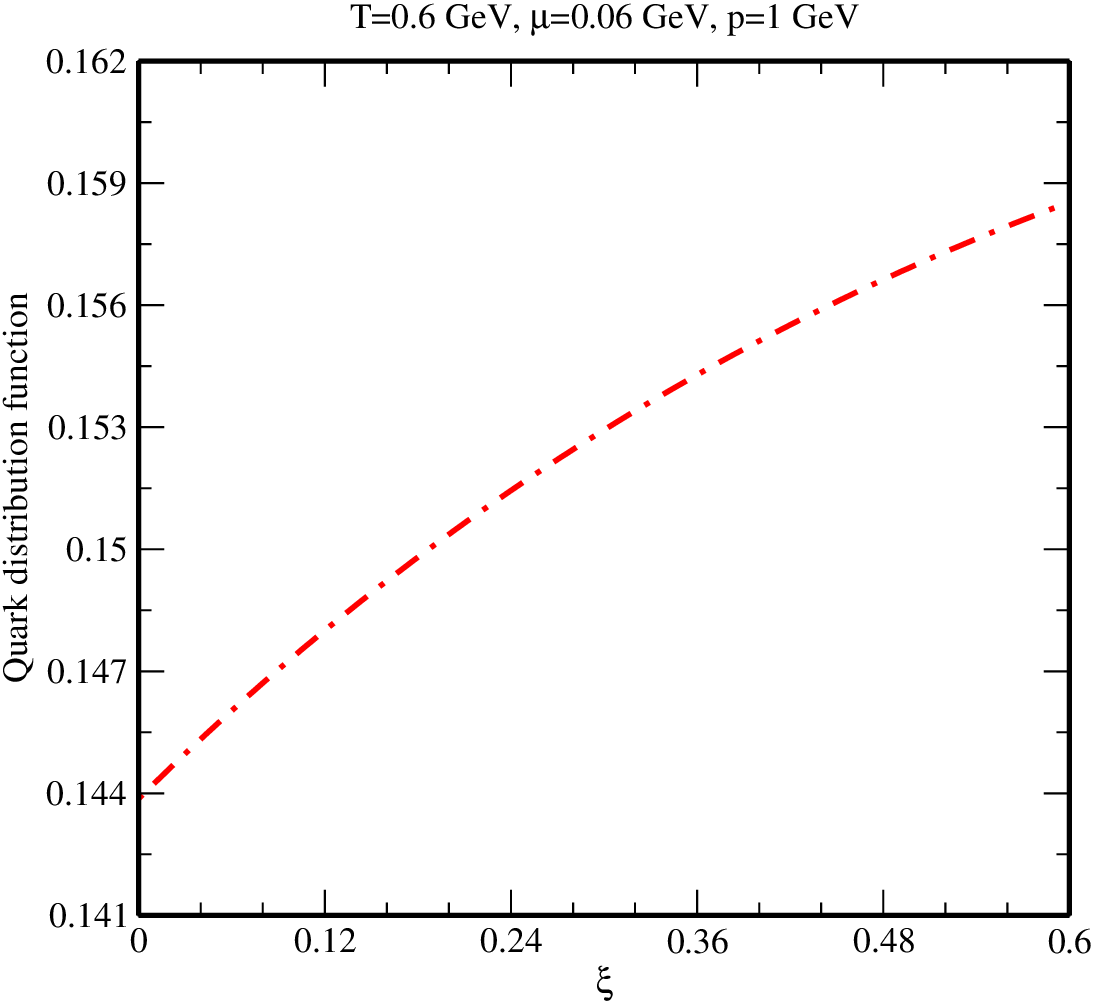} \\
a & b
\end{tabular}
\caption{Variations of the quark distribution function with anisotropy parameter (a) at $T=0.16$ GeV and (b) at $T=0.6$ GeV for a fixed chemical potential and a fixed momentum.}\label{Fig.dftu}
\end{center}
\end{figure}

\begin{figure}[]
\begin{center}
\begin{tabular}{c c}
\includegraphics[width=7.4cm]{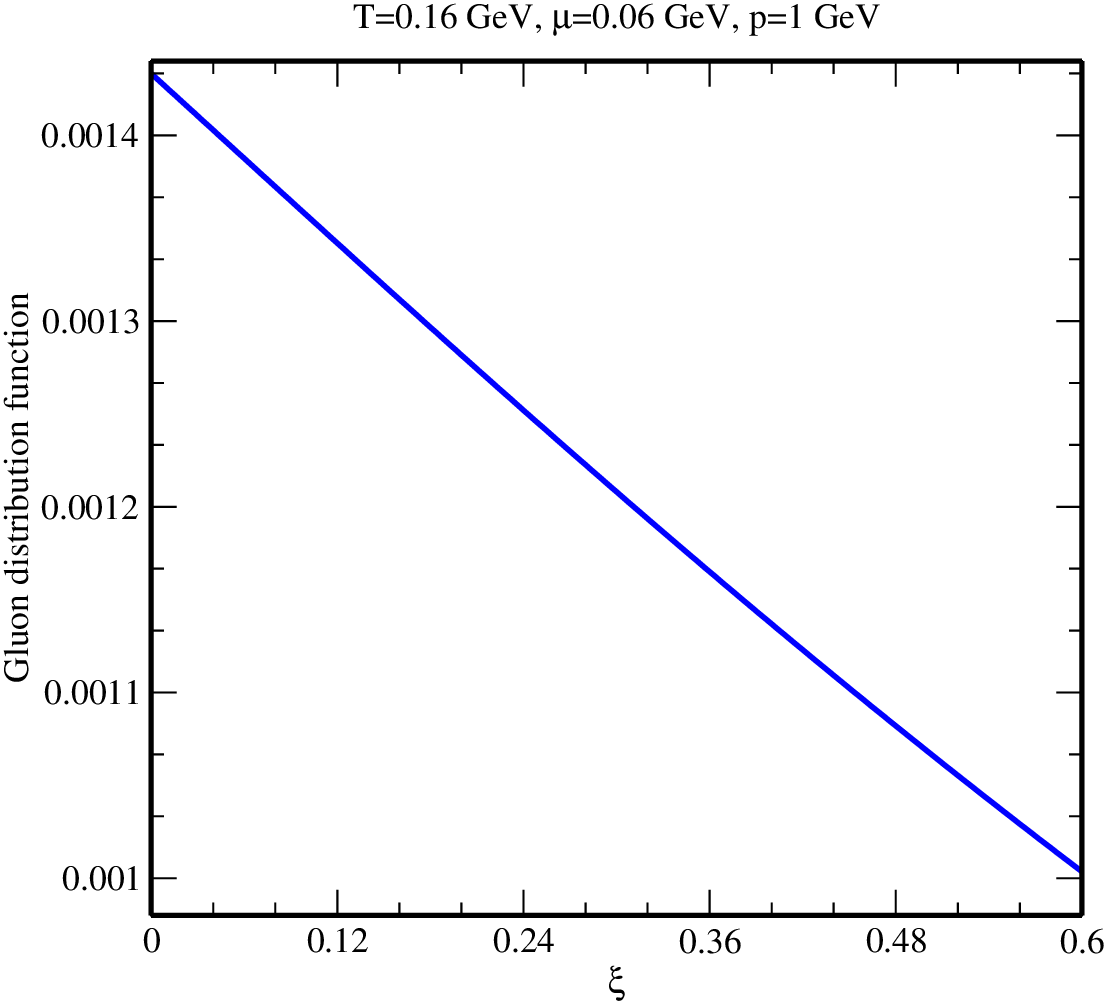}&
\hspace{0.73 cm}
\includegraphics[width=7.4cm]{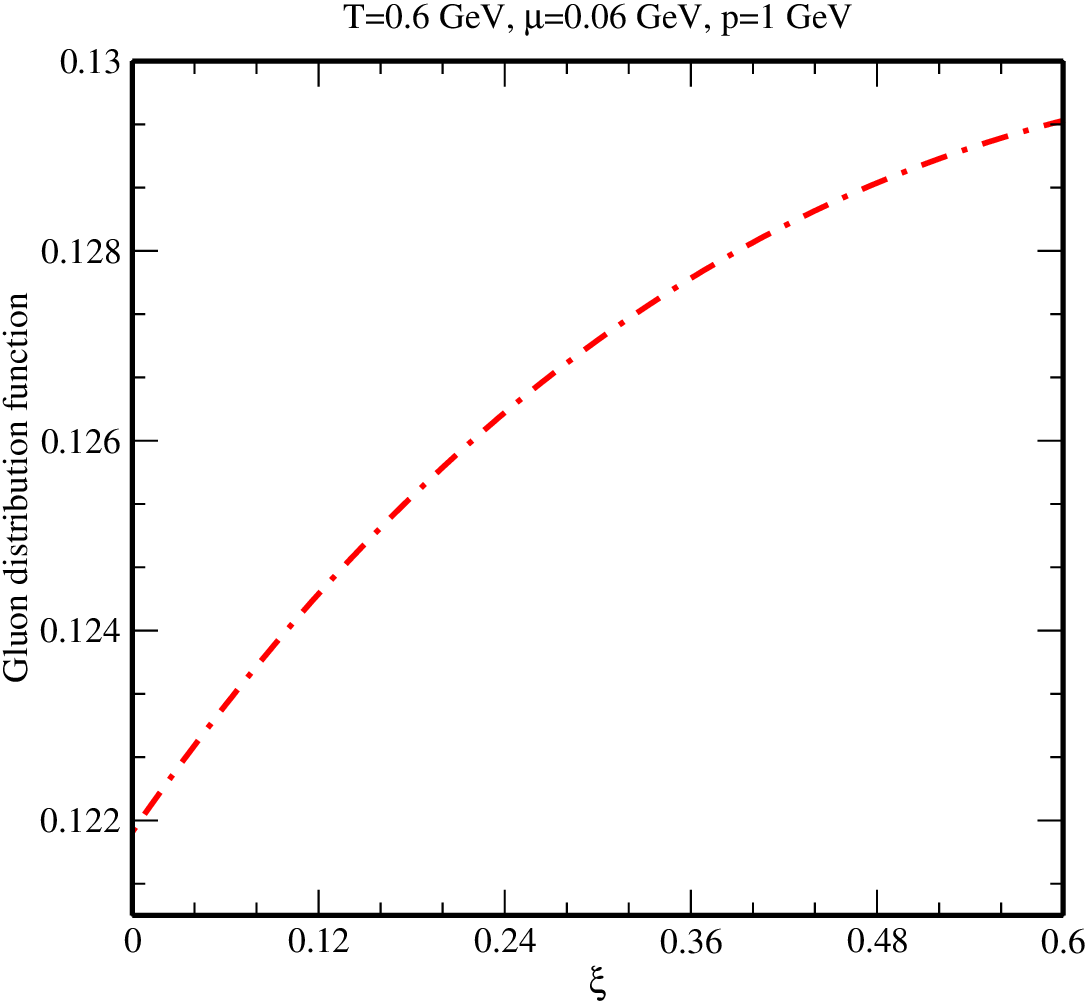} \\
a & b
\end{tabular}
\caption{Variations of the gluon distribution function with anisotropy parameter (a) at $T=0.16$ GeV and (b) at $T=0.6$ GeV for a fixed chemical potential and a fixed momentum.}\label{Fig.dftg}
\end{center}
\end{figure}

We first examine the dependence of the quasiparticle masses of quark and gluon on the anisotropy parameter at low and high temperatures for a fixed chemical potential. As per figures \ref{Fig.mq} and \ref{Fig.mg}, increasing the expansion-induced anisotropy leads to a monotonic enhancement of the quasiparticle masses, resulting in masses larger than those in the isotropic medium. This anisotropic enhancement is relatively weak at low temperatures, as seen in figures \ref{Fig.mq}a and \ref{Fig.mg}a, but becomes increasingly pronounced at higher temperatures, as illustrated in figures \ref{Fig.mq}b and \ref{Fig.mg}b. The anisotropy-induced modification of the quasiparticle masses consequently alters the partonic dispersion relations and, hence, the corresponding distribution functions also get modified. In the isotropic medium, the distribution functions depend on the temperature- and chemical potential-dependent quasiparticle masses as given in equations \eqref{Q.P.M.(Quark mass)} and \eqref{Q.P.M.(Gluon mass)}, whereas in the expansion-induced anisotropic medium they involve quasiparticle masses which additionally depend on the anisotropy parameter, as specified in equations \eqref{Q.P.M.Q.(Anisotropic medium)} and \eqref{Q.P.M.G.(Anisotropic medium)}. To illustrate this behavior, figures \ref{Fig.dftu} and \ref{Fig.dftg} show the quark and gluon distribution functions, respectively, as functions of the anisotropy parameter for fixed values of the temperature, chemical potential and momentum. At low temperatures, both quark and gluon distribution functions decrease gradually with increasing anisotropy parameter, as shown in figures \ref{Fig.dftu}a and \ref{Fig.dftg}a. In contrast, at high temperatures, these distribution functions increase more noticeably with the anisotropy parameter, as observed in figures \ref{Fig.dftu}b and \ref{Fig.dftg}b. The combined behavior of the quasiparticle masses and distribution functions demonstrates how expansion-induced anisotropy modifies the microscopic properties of baryon asymmetric hot QCD medium. These modifications are consequently transmitted to the shear and bulk viscosities and their associated observables. 

\section{Results and discussions}
\begin{figure}[]
\begin{center}
\begin{tabular}{c c}
\includegraphics[width=7.4cm]{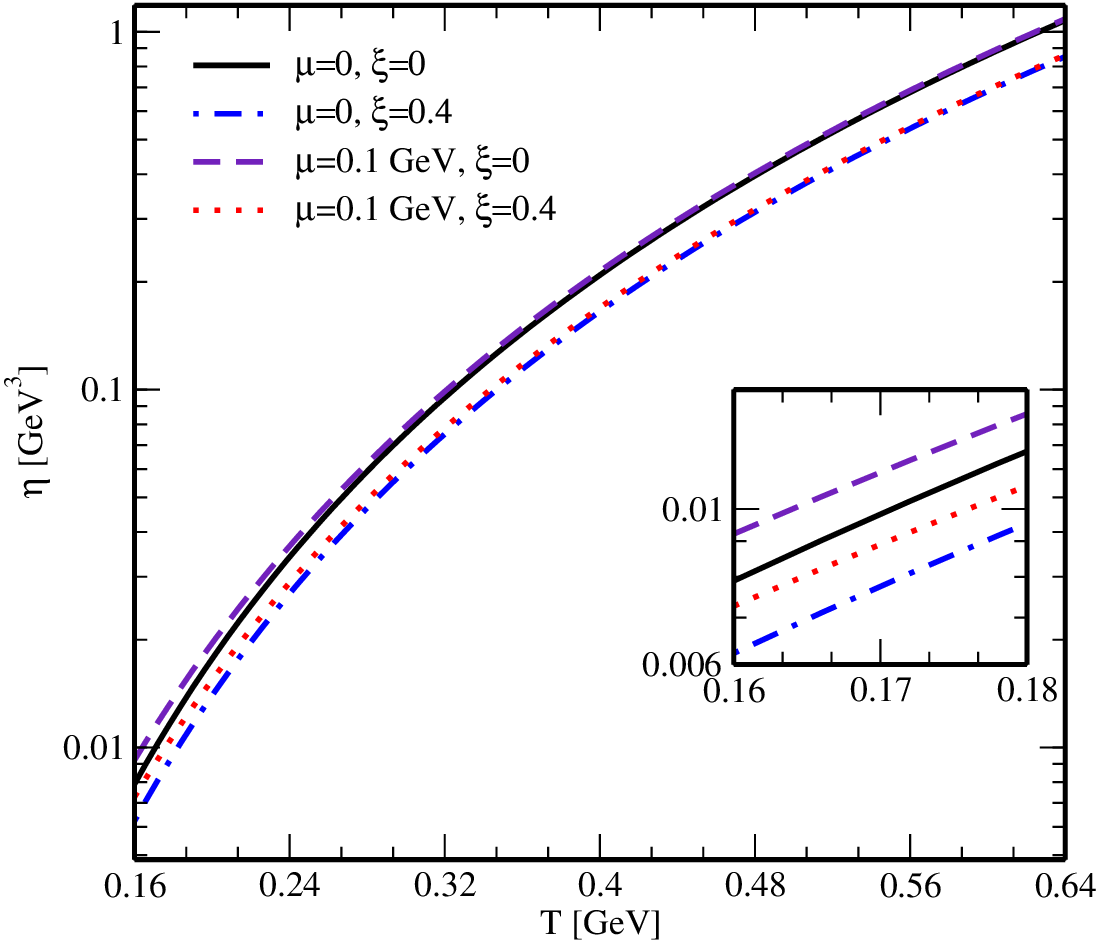}&
\hspace{0.73 cm}
\includegraphics[width=7.4cm]{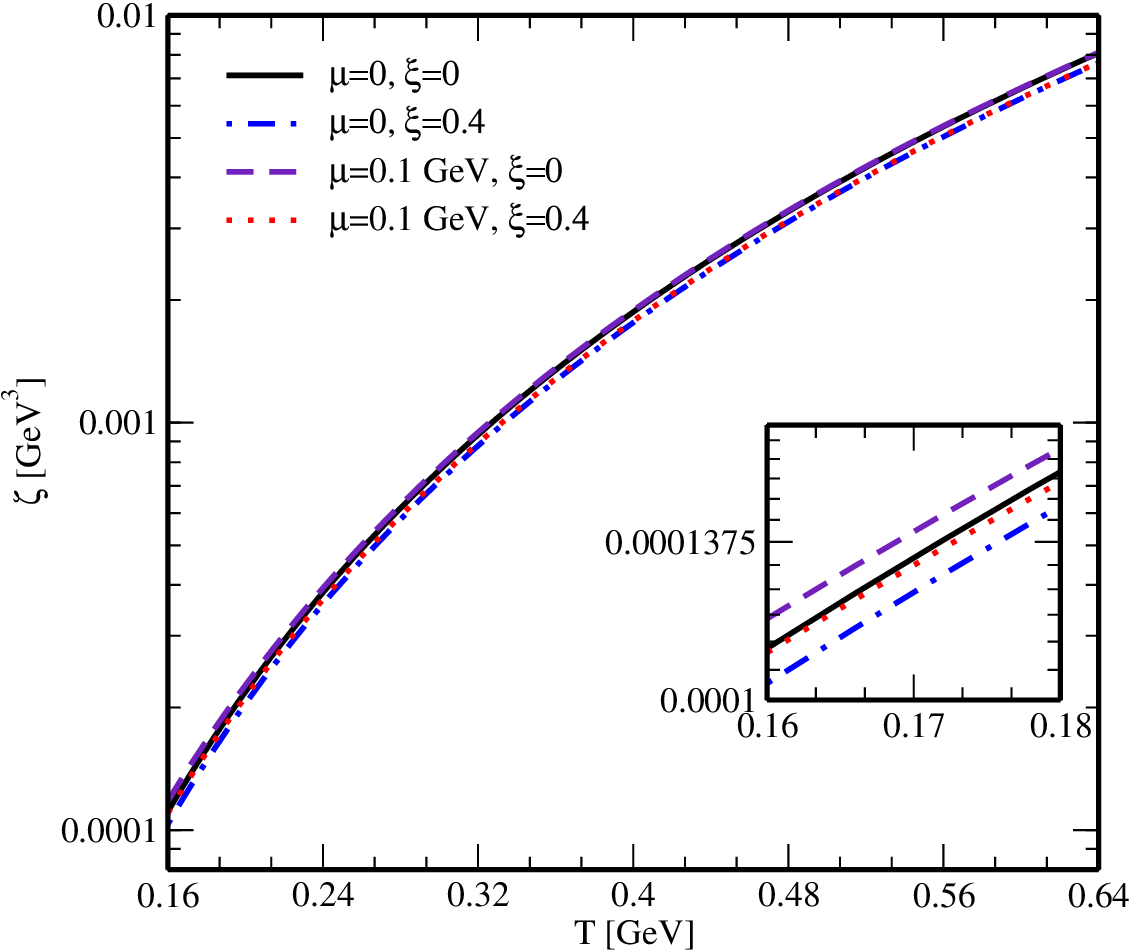} \\
a & b
\end{tabular}
\caption{Variations of (a) the shear viscosity and (b) the bulk viscosity with temperature for different values of anisotropy parameter and chemical potential.}\label{Fig.1}
\end{center}
\end{figure}

Figures \ref{Fig.1}a and \ref{Fig.1}b present the temperature dependence of the shear viscosity ($\eta$) and the bulk viscosity ($\zeta$), respectively, for different values of the anisotropy parameter and chemical potential. Both viscosities increase monotonically with temperature across all considered cases. Introduction of expansion-induced anisotropy, however, leads to a reduction in both $\eta$ and $\zeta$ relative to the corresponding isotropic results, for both baryonless matter and baryon asymmetric matter. This indicates that the momentum transfer within the medium and the fluctuations in local pressure are suppressed as the degree of anisotropy increases. The anisotropy dependence of $\eta$ and $\zeta$ can be understood from equations \eqref{aniso.eta1} and \eqref{aniso.zeta1}, where the anisotropic contributions $\eta^\xi$ and $\zeta^\xi$ depend explicitly on the anisotropy parameter $\xi$ and implicitly on $\xi$ through the modified quasiparticle dispersion relations and distribution functions. The anisotropy-induced increase in the quasiparticle masses, together with the corresponding modifications of the distribution functions, enhances the magnitude of the negative anisotropic contributions to the viscosities. Although the distribution functions exhibit a mild decrease with increasing $\xi$ at low temperatures, their overall behavior at higher temperatures, together with the anisotropy-dependent quasiparticle masses, leads to increasingly negative contributions from $\eta^\xi$ and $\zeta^\xi$. Consequently, the total shear and bulk viscosities decrease with increasing anisotropy. In contrast, baryon asymmetric matter exhibits larger values of both $\eta$ and $\zeta$ than baryonless matter, irrespective of whether the medium is isotropic or anisotropic. The effect of finite chemical potential is particularly pronounced at low temperatures and becomes comparatively weaker at higher temperatures. This enhancement is primarily associated with the increased parton number densities in the presence of finite baryon asymmetry, which strengthens the momentum transport and the dissipative response to local pressure variations. Thus, baryon asymmetry enhances both shear and bulk viscous effects, while expansion-induced anisotropy acts to suppress them, with these competing effects persisting in both isotropic and anisotropic QCD media. 

\begin{figure}[]
\begin{center}
\begin{tabular}{c c}
\includegraphics[width=7.4cm]{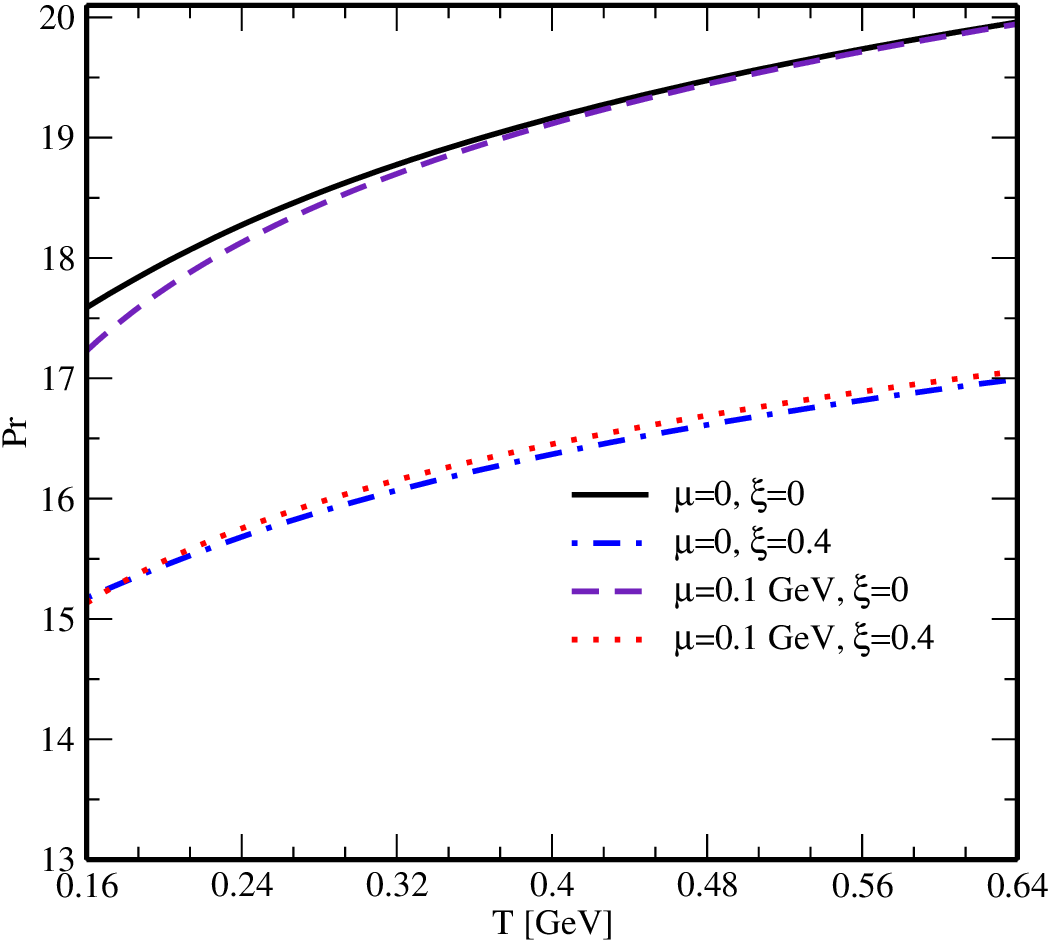}&
\hspace{0.73 cm}
\includegraphics[width=7.4cm]{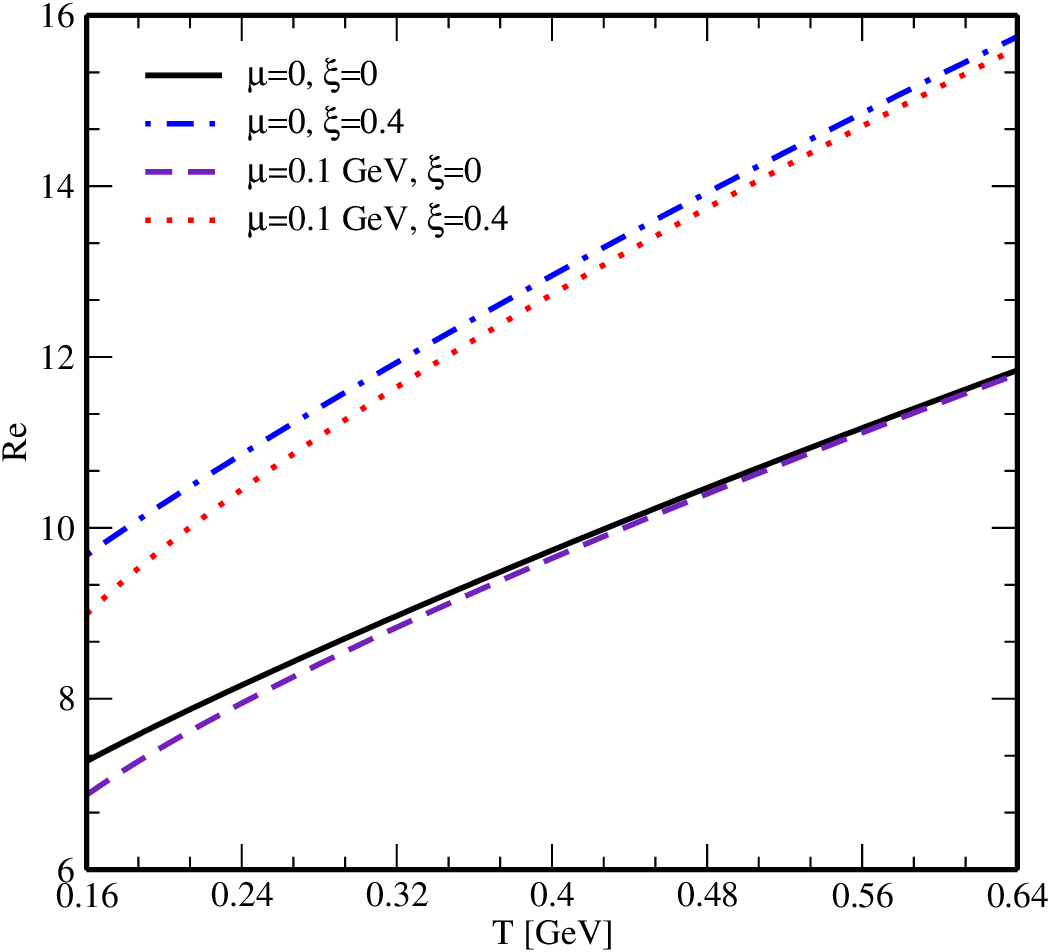} \\
a & b
\end{tabular}
\caption{Variations of (a) the Prandtl number and (b) the Reynolds number with temperature for different values of anisotropy parameter and chemical potential.}\label{Fig.2}
\end{center}
\end{figure}

\begin{figure}[]
\begin{center}
\begin{tabular}{c c}
\includegraphics[width=7.4cm]{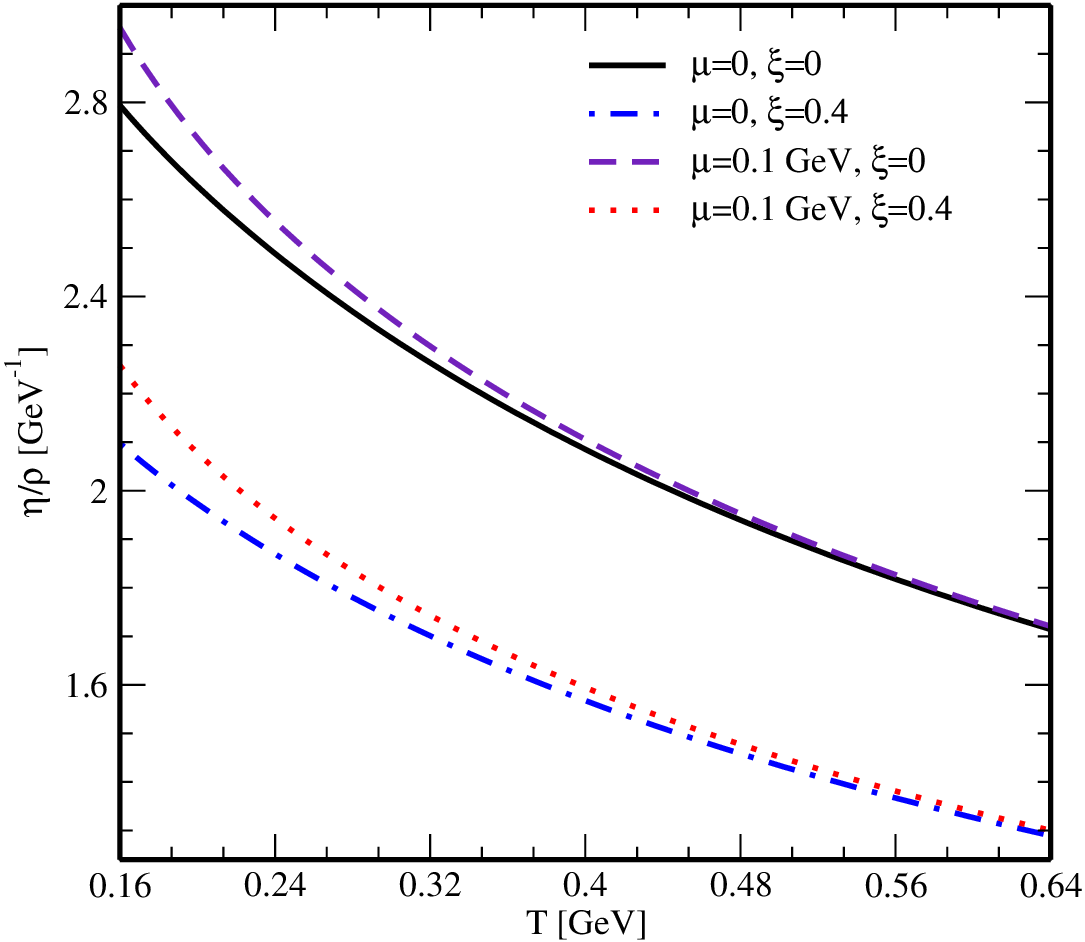}&
\hspace{0.73 cm}
\includegraphics[width=7.4cm]{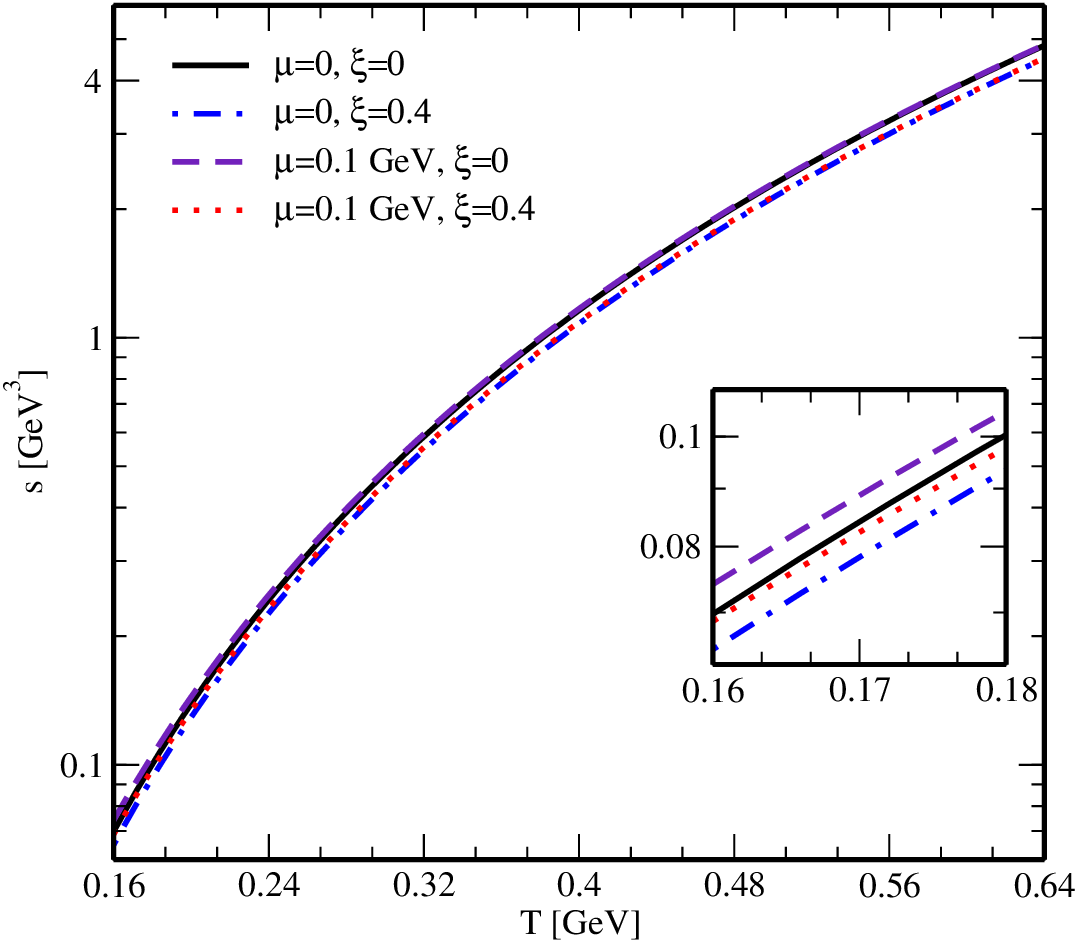} \\
a & b
\end{tabular}
\caption{Variations of (a) the kinematic viscosity and (b) the entropy density with temperature for different values of anisotropy parameter and chemical potential.}\label{Fig.3}
\end{center}
\end{figure}

\begin{figure}[]
\begin{center}
\begin{tabular}{c c}
\includegraphics[width=7.4cm]{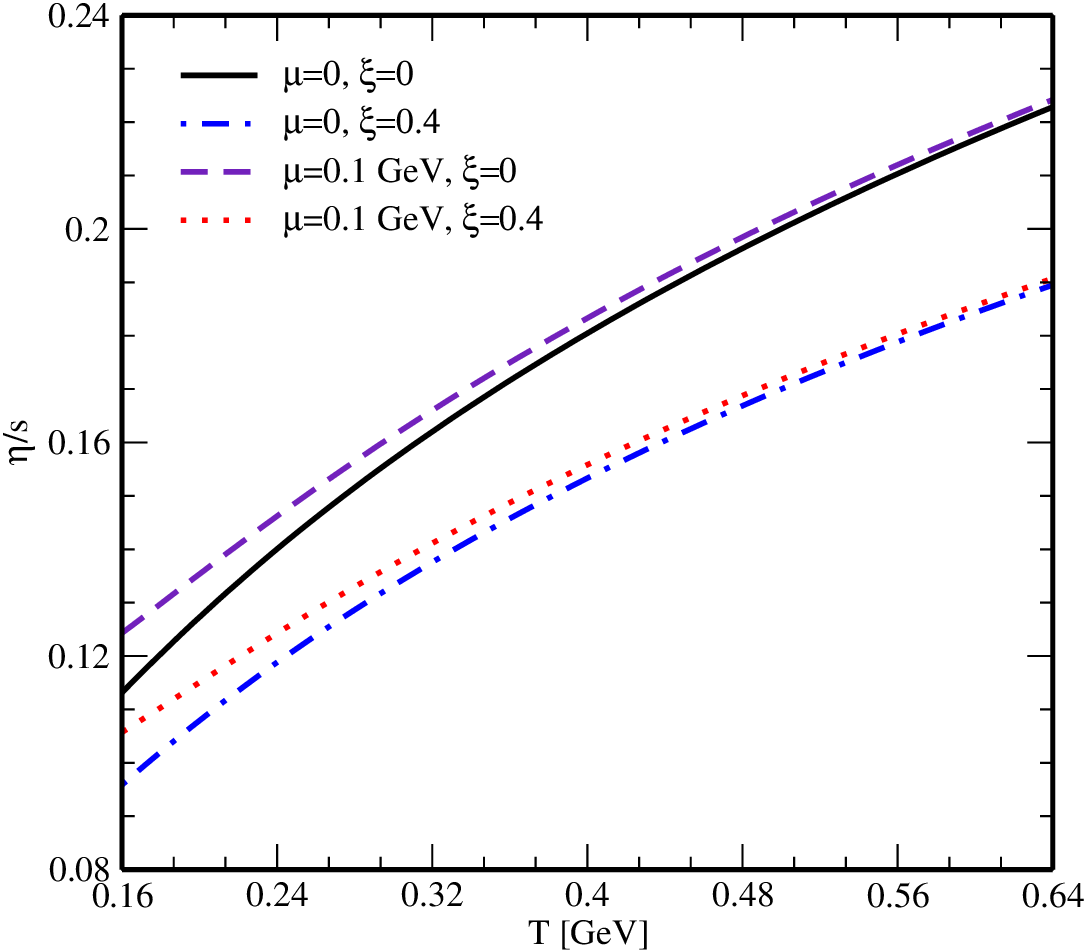}&
\hspace{0.73 cm}
\includegraphics[width=7.4cm]{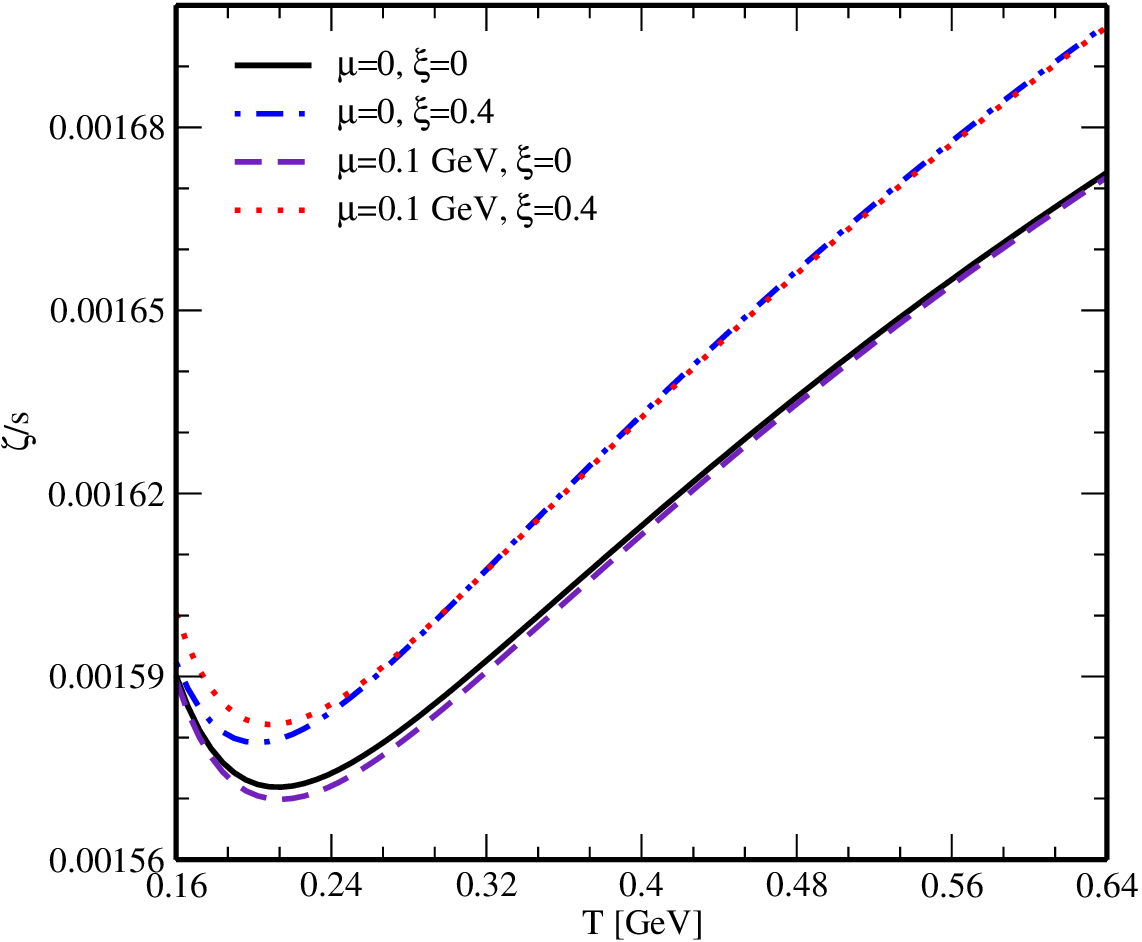} \\
a & b
\end{tabular}
\caption{Variations of (a) the specific shear viscosity and (b) the specific bulk viscosity with temperature for different values of anisotropy parameter and chemical potential.}\label{Fig.4}
\end{center}
\end{figure}

Figure \ref{Fig.2}a shows the temperature dependence of the Prandtl number for four cases, such as baryonless isotropic, baryonless anisotropic, baryon asymmetric isotropic and baryon asymmetric anisotropic media. For 
all four cases, the Prandtl number increases monotonically with temperature and remains above unity throughout the considered range. This indicates that the momentum diffusion is more efficient than the thermal diffusion, implying that momentum transport provides the dominant contribution to the sound attenuation in these media. For the baryonless medium, the introduction of expansion-induced anisotropy leads to a noticeable reduction in the Prandtl number relative to the isotropic case. A similar, although comparatively smaller, reduction is observed when anisotropy is introduced into baryon asymmetric medium. Thus, the expansion-induced anisotropy weakens the relative dominance of momentum diffusion over thermal diffusion, thereby modifying the balance between the two dissipative transport mechanisms relevant to the sound attenuation. The effect of baryon asymmetry is comparatively modest and depends on whether the medium is isotropic or anisotropic. At finite baryon asymmetry, the Prandtl number decreases slightly in the isotropic medium, indicating a small reduction in the relative contribution of momentum diffusion to the sound attenuation. However, in the anisotropic medium, finite baryon asymmetry produces a slight increase in the Prandtl number, corresponding to a modest enhancement of the relative importance of momentum diffusion. These contrasting responses demonstrate that the baryon asymmetry modifies the interplay between momentum diffusion and thermal diffusion differently in isotropic and anisotropic media. Although the Prandtl number exhibits the same qualitative temperature dependence in all cases, the effects of finite chemical potential and anisotropy are manifested primarily through changes in its magnitude. Among the four media considered, the baryonless isotropic medium has the highest value of the Prandtl number, whereas the baryonless anisotropic medium has the lowest value of the Prandtl number. Accordingly, the relative dominance of momentum diffusion over thermal diffusion is strongest in the former case and weakest in the latter case. Overall, both expansion-induced anisotropy and baryon asymmetry modify the relative contributions of momentum diffusion and thermal diffusion and, consequently, leave discernible imprints on the sound attenuation in hot QCD matter. 

Figure \ref{Fig.2}b presents the temperature dependence of the Reynolds number for the four media considered above. The Reynolds number increases with temperature in all cases, but remains close to unity and does not become substantially larger than one over the considered temperature range. This indicates that the viscous effects remain important and the flow remains predominantly laminar rather than developing into a strongly turbulent flow. The observed increase of the Reynolds number with the temperature can be attributed to the corresponding decrease in the kinematic viscosity, as shown in figure \ref{Fig.3}a. The introduction of expansion-induced anisotropy increases the Reynolds number in both baryonless and baryon asymmetric media. In contrast, finite baryon asymmetry reduces the Reynolds number relative to the corresponding baryonless cases, with this effect being more pronounced at lower temperatures. Comparing the two effects, we have observed that the modification induced by expansion-driven anisotropy is more substantial than that associated with baryon asymmetry. Among the four media, the baryonless anisotropic medium exhibits the highest value of the Reynolds number, whereas the baryon asymmetric isotropic medium estimates the lowest value of the Reynolds number. The corresponding differences reflect variations in the relative importance of inertial and viscous effects. A larger Reynolds number indicates comparatively weaker role of viscous dissipation, while its smaller value signifies a stronger viscous influence. Nevertheless, since the Reynolds number remains close to unity in all cases, viscous effects remain relevant across the entire temperature range considered. 

Figure \ref{Fig.3}b illustrates the temperature dependence of the entropy density for different values of anisotropy parameter and chemical potential. The entropy density increases with the temperature for all considered cases. Among the four media, it attains its largest value in the baryon asymmetric isotropic medium and its smallest value in the baryonless anisotropic medium. Since the entropy density characterizes the number of accessible microscopic states, these results indicate a greater number of accessible microstates in the former case and a comparatively smaller number of accessible microstates in the latter case. Thus, finite baryon asymmetry enhances the disorderliness of the medium, whereas expansion-induced anisotropy tends to reduce it. The combined behavior of the entropy density with that of the shear and bulk viscosities provides the basis for understanding the temperature dependence of the specific shear ($\eta/s$) and specific bulk ($\zeta/s$) viscosities. Figures \ref{Fig.4}a and \ref{Fig.4}b show the temperature dependence of $\eta/s$ and $\zeta/s$, respectively, for the four media considered above. As per figure \ref{Fig.4}a, the ratio $\eta/s$ increases with the temperature in all cases. The introduction of expansion-induced anisotropy reduces $\eta/s$ in both baryonless and baryon asymmetric media, whereas finite baryon asymmetry enhances $\eta/s$ for both isotropic and anisotropic media. Among the four cases, the largest value of $\eta/s$ is obtained for the baryon asymmetric isotropic medium, while the smallest value occurs in the baryonless anisotropic medium. In the latter case, $\eta/s$ remains closest to the conjectured lower bound $1/(4\pi)$ from the AdS/CFT correspondence, indicating behavior closer to that associated with a nearly perfect fluid. The ratio $\zeta/s$, as shown in figure \ref{Fig.4}b, also exhibits an overall increasing tendency with the temperature, although its temperature dependence is nonmonotonic in certain parameter ranges. In particular, a distinct minimum or dip appears for specific values of temperature, anisotropy parameter and chemical potential. The value of $\zeta/s$ at this minimum differs among the four media, with the smallest value for the baryon asymmetric isotropic medium and the largest value for the baryon asymmetric anisotropic medium. Since a vanishing bulk viscosity is associated with conformal symmetry, the magnitude of $\zeta/s$ provides an indication of the degree of conformal symmetry in the medium. The relatively smaller $\zeta/s$ for the baryon asymmetric isotropic medium therefore indicates that this medium is closer to conformal symmetry, whereas the larger value in the presence of both anisotropy and baryon asymmetry reflects a stronger deviation from conformal behavior. Overall, the combined observations of $\eta/s$ and $\zeta/s$ demonstrate that the expansion-induced anisotropy and the baryon asymmetry influence the dissipative and conformal properties of hot QCD medium. 

\section{Conclusions}
In this work, we studied the influence of expansion-induced anisotropy on the viscous properties and observables of baryon asymmetric hot QCD medium. In particular, we examined the shear viscosity, 
bulk viscosity, the Prandtl number, the Reynolds number, specific shear viscosity and specific bulk 
viscosity. The shear and bulk viscosities were evaluated by solving the relativistic Boltzmann transport equation in the relaxation time approximation within the kinetic theory framework. The effects of partonic interactions were incorporated through a quasiparticle description, in which the quark and gluon distribution functions and their temperature-, chemical potential- and anisotropy-dependent effective masses accounted for the properties of the medium. We found that the quasiparticle masses of quarks and gluons increased with the anisotropy parameter, reflecting the modification of the partonic dispersion relations induced by expansion-driven momentum anisotropy. These modifications, together with the corresponding changes in the distribution functions, resulted in a reduction of both shear and bulk viscosities with increasing anisotropy. In contrast, finite baryon asymmetry enhanced both viscosities relative to their baryonless counterparts, with its effect being more pronounced at lower temperatures. Consequently, expansion-induced anisotropy suppressed momentum transfer and dissipative response to local pressure variations, whereas baryon asymmetry enhanced them. Thus, expansion-induced anisotropy and baryon asymmetry produced opposite effects on the viscous response of the medium. The dimensionless observables considered in this work also exhibited a clear sensitivity to the medium conditions. The Prandtl number remained above unity and increased with temperature, indicating the dominance of momentum diffusion over thermal diffusion in the sound attenuation process. Expansion-induced anisotropy reduced the Prandtl number, whereas it increased the Reynolds number. Nevertheless, the Reynolds number did not significantly exceed unity, indicating that viscous effect continued to play an important role in the flow dynamics. The specific shear viscosity increased with temperature and was reduced by anisotropy but enhanced by baryon asymmetry. On the other hand, the specific bulk viscosity exhibited a nonmonotonic temperature dependence in certain parameter regions, reflecting the sensitivity of conformal symmetry to the thermodynamic and anisotropic properties of the medium. Overall, our results demonstrated that the expansion-induced anisotropy and baryon asymmetry left distinct imprints on the viscous properties of hot QCD medium. The effects of expansion-induced anisotropy were generally more pronounced for the Prandtl number, the Reynolds number, specific shear viscosity and specific bulk viscosity than those associated with baryon asymmetry. These findings highlighted the importance of accounting for both expansion-induced anisotropy and baryon density when characterizing the viscous properties and related observables of hot QCD medium produced under extreme conditions. 

\section{Acknowledgments}
One of the authors (S. R.) acknowledges financial support from ANID Fondecyt Postdoctoral Grant 3240349 
for this work. N. N. acknowledges support from ANID (Chile) FONDECYT Iniciaci\'on Grant No. 11230879.

\end{document}